\newif\ifdraft
\newif\iffull
\newif\ifcomment
\newif\iflatexdiff
\newif\ifbibtex
\newif\ifpreprint
\newif\ifbanner
\newif\ifsupp
\preprinttrue    
\fulltrue        

\def\dvers{v10}

\ifpreprint

\fi
\def\dtitle{Freeze-out radii extracted from three-pion cumulants \\
in \pp, \pPb\ and \PbPb\ collisions at the LHC} 
\def\stitle{Freeze-out radii in \pp, \pPb\ and \PbPb\ from three-pion cumulants} 
\ifpreprint
\documentclass[ALICE,manyauthors,12pt]{cernphprep}
\usepackage[comma,square,numbers,sort&compress]{natbib}
\usepackage{hyperref}
\else
\documentclass[final,3p,12pt]{elsarticle}
\biboptions{comma,square,numbers,sort&compress}
\usepackage{hyperref}
\fi
\usepackage{graphicx}  
\usepackage{dcolumn}   
\usepackage{bm}        
\usepackage{amssymb}   
\usepackage{amsfonts}
\usepackage{graphics}
\usepackage{grffile}   
\usepackage{epsfig}
\usepackage{units}
\usepackage[usenames]{color}
\usepackage[normalem]{ulem} 
\usepackage{color}
\usepackage[utf8]{inputenc}
\usepackage[T1]{fontenc}
\usepackage{subfigure}
\iflatexdiff
\RequirePackage{color}\definecolor{RED}{rgb}{1,0,0}\definecolor{BLUE}{rgb}{0,0,1}

\fi

\ifpreprint
\else
 
\fi

\newcommand{\fc}           {\ensuremath{f_{\rm c}}}
\newcommand{\kT}           {\ensuremath{k_{\rm T}}}
\newcommand{\KTThree}      {\ensuremath{K_{\rm T,3}}}
\newcommand{\qinv}         {\ensuremath{q}}
\newcommand{\Rinv}         {\ensuremath{R_{\rm inv}}}
\newcommand{\Rinvt}        {\ensuremath{R_{\rm inv,3}}}
\newcommand{\ls}           {\ensuremath{\lambda_{\rm e}}}
\newcommand{\lst}          {\ensuremath{\lambda_{\rm e,3}}}

\newcommand{\dEdx}         {d$E$/d$x$}
\newcommand{\Nch}          {\ensuremath{\left< N_{\rm ch} \right>}}
\newcommand{\Nrecp}        {\ensuremath{N^{\rm rec}_{\rm pions}}}

\newcommand{\avNp}         {\ensuremath{\left< N_{\rm pions} \right>}}

\newcommand{\pp}           {pp}
\newcommand{\ppbar}        {\mbox{$\mathrm{p\overline{p}}$}}
\newcommand{\PbPb}         {\mbox{Pb--Pb}}

\newcommand{\pPb}          {\mbox{p--Pb}}

\newcommand{\pt}           {\ensuremath{p_{\mathrm{T}}}{ }}

\newcommand{\snn}          {\ensuremath{\sqrt{s_{\mathrm{NN}}}}}

\newcommand{\Fig}[1]       {Fig.~\ref{#1}}

\newcommand{\Ref}[1]       {Ref.~\cite{#1}}

\newcommand{\red}[1]       {\textcolor{red}{#1}}

\newcommand{\warn}[1]      {{\small\textbf{\red{(!}\footnote{\textbf{\red{(!)}}~#1}\red{)}}}\marginpar{\textbf{\red{---}}}}

\newcommand{\arxiv}[1]     {\href{http://www.arxiv.org/abs/#1}{\mbox{arXiv:#1}}}

\newcommand{\com}[1]       {}

\iflatexdiff

\renewcommand{\xout}[1]    {\textcolor{red}{\sout{#1}}}
 
\newcommand{\old}[1]       {{\textcolor{red}{\sout{#1}}}}

\else

\renewcommand{\xout}[1]    {}
\newcommand{\old}[1]       {\relax}

\fi

\graphicspath{{./img/}}
\ifdraft
\usepackage{lineno}
\linenumbers
\modulolinenumbers[1]
\usepackage{fancyhdr}
\ifbanner
\fi

\else
\renewcommand{\warn}[1]{}

\fi
\begin{document}
\newlength{\figlen}
\setlength{\figlen}{\linewidth}
\ifpreprint
\setlength{\figlen}{0.95\linewidth}
\begin{titlepage}
\PHyear{2014}
\PHnumber{062}                        
\PHdate{27 March}                     
\title{\dtitle}
\ShortTitle{\stitle}
\Collaboration{ALICE Collaboration%
         \thanks{See Appendix~\ref{app:collab} for the list of collaboration members}}
\ShortAuthor{ALICE Collaboration} 
\ifdraft
\begin{center}
\ifbanner
 \today\\ \color{red}DRAFT \dvers\ \hspace{0.3cm} \$Revision: 1735 $\color{white}:$\$\color{black}\vspace{0.3cm}
\else
 \today\\ \color{red}DRAFT \dvers\ \color{black}\vspace{0.3cm}
\fi
\end{center}
\fi
\else
\begin{frontmatter}
\title{\dtitle}
\iffull
\input{Alice_Authorlist_2014-Mar-21-PLB}
\else
\ifdraft
\author{ALICE Collaboration \\ \vspace{0.3cm} 
\today\\ \color{red}DRAFT \dvers\ \hspace{0.3cm} (\$Revision: 1735 $\color{white}:$\$)\color{black}}
\else
\author{ALICE Collaboration}
\fi
\fi
\fi
\begin{abstract}
In high-energy collisions, the spatio-temporal size of the particle production region can be measured 
using the Bose-Einstein correlations of identical bosons at low relative momentum.  
The source radii are typically extracted using two-pion correlations, and characterize the system at the 
last stage of interaction, called kinetic freeze-out.
In low-multiplicity collisions, unlike in high-multiplicity collisions,
two-pion correlations are substantially altered by background correlations, e.g.\ mini-jets.
Such correlations can be suppressed using three-pion cumulant correlations.
We present the first measurements of the size of the system at freeze-out extracted from three-pion cumulant correlations 
in \pp, \pPb\ and \PbPb\ collisions at the LHC with ALICE.
At similar multiplicity, the invariant radii extracted in \pPb\ collisions are found to be 5--15\% larger than those in pp, 
while those in \PbPb\ are 35--55\% larger than those in p--Pb.
Our measurements disfavor models which incorporate substantially stronger collective expansion in \pPb\ 
as compared to \pp\ collisions at similar multiplicity.

\ifdraft 
\ifpreprint
\end{abstract}
\end{titlepage}
\else
\end{abstract}
\end{frontmatter}
\newpage
\fi
\fi
\ifdraft
\thispagestyle{fancyplain}
\else
\end{abstract}
\ifpreprint
\end{titlepage}
\else
\end{frontmatter}
\fi
\fi
\setcounter{page}{2}


\section{Introduction}
\label{sec:intro}
The role of initial and final-state effects in interpreting differences between \PbPb\ and \pp\ collisions is expected 
to be clarified with \pPb\ collisions~\cite{Salgado:2011wc}.  
However, the results obtained from \pPb\ collisions at $\snn=5.02$~TeV~\cite{CMS:2012qk,Abelev:2012ola,Aad:2012gla,Aad:2013fja,Chatrchyan:2013nka,Chatrchyan:2013eya,Abelev:2013haa,Abelev:2013wsa,Abelev:2013bla} have not been conclusive since they can be explained assuming either a hydrodynamic phase during the evolution of the system~\cite{Bozek:2012gr,Bozek:2013uha,Qin:2013bha} or the formation of a Color Glass Condensate~(CGC) in the initial state~\cite{Dusling:2012wy,Dusling:2013oia}.

As in \PbPb\ collisions, the presence of a hydrodynamic phase in high-multiplicity \pPb\ collisions is expected to lead 
to a factor of $1.5$--$2$ larger freeze-out radii than in \pp\ collisions at similar multiplicity~\cite{Bozek:2013df}. 
In contrast, a CGC initial state model (IP-Glasma), without a hydrodynamic phase, predicts similar freeze-out radii in \pPb\ and \pp\ collisions~\cite{Bzdak:2013zma}. 
A measurement of the freeze-out radii in the two systems will thus lead to additional experimental constraints on the interpretation of the \pPb\ data.

The extraction of freeze-out radii can be achieved using identical boson correlations at low relative momentum,
which are dominated by quantum statistics (QS) and final-state Coulomb and strong interactions (FSIs). 
Both FSIs and QS correlations encode information about the femtoscopic space-time structure of the particle emitting 
source at kinetic freeze-out~\cite{Goldhaber:1960sf,Kopylov:1975rp,Lednicky:2005tb}.
The calculation of FSI correlations allows for the isolation of QS correlations.
Typically, two-pion QS correlations are used to extract the characteristic radius of the 
source~\cite{Adams:2004yc,Lisa:2005dd,Khachatryan:2010un,Aamodt:2010jj,Khachatryan:2011hi,Aamodt:2011kd,Aamodt:2011mr}.
However, higher-order QS correlations can be used as well~\cite{Neumeister:1991bq,Bearden:2001ea,Ackerstaff:1998py,Abreu:1995sq,Achard:2002ja}.
The novel features of higher-order QS correlations are extracted using the cumulant for which all lower order correlations are removed~\cite{Heinz:1997mr,Heinz:2004pv}.
The maximum of the three-pion cumulant QS correlation is a factor of two larger than for two-pion QS correlations~\cite{Weiner:1988jy,Andreev:1992pu,Heinz:1997mr,Heinz:2004pv}.
In addition to the increased signal, three-pion cumulants also remove contributions from two-particle 
background correlations unrelated to QS~(e.g.\ from mini-jets~\cite{Aamodt:2010jj,Aamodt:2011kd}).
The combined effect of an increased signal and decreased background is advantageous in low multiplicity systems where a substantial background exists.  

In this letter, we present measurements of freeze-out radii extracted using three-pion cumulant QS correlations.  
The invariant radii are extracted in intervals of multiplicity and triplet momentum in \pp~($\sqrt{s}=7$ TeV), \pPb~($\snn=5.02$ TeV) and 
\PbPb~($\snn=2.76$ TeV) which allows for a comparison of the various systems. The radii extracted from three-pion cumulants are 
also compared to those from two-pion correlations.

The letter is organized into 5 remaining sections:
Section \ref{sec:setup} explains the experimental setup and event selection.
Section \ref{sec:selection} describes the identification of pions, as well as the measurement of the event multiplicity.
Section \ref{sec:analysis} explains the three-pion cumulant analysis technique used to extract the source radii.
Section \ref{sec:results} presents the measured source radii.
Finally, Section \ref{sec:summary} summarizes the results reported in the letter.
\ifsupp
Section \ref{app:supp} lists a collection of supplemental figures.
\fi

\section{Experimental setup and event selection}
\label{sec:setup}
Data from \pp, \pPb\, and \PbPb\ collisions at the LHC recorded with ALICE~\cite{Aamodt:2008zz} are analyzed.   
The data for \pp\ collisions were taken during the 2010 \pp\ run at $\sqrt{s}=7$~TeV,
for \pPb\ collisions during the 2013 run at $\snn=5.02$~TeV,
and for \PbPb\ during the 2010 and 2011 runs at $\snn=2.76$~TeV.
For \pPb, the proton beam energy was \unit[4]{TeV} while for the lead beam it was \unit[1.58]{TeV} per nucleon.
Thus, the nucleon--nucleon center-of-mass system moved with respect to the ALICE laboratory system 
with a rapidity of $-0.465$, i.e.\ in the direction of the proton beam.
The pseudorapidity in the laboratory system is denoted with $\eta$ throughout this letter, which for
the \pp\ and \PbPb\ systems coincides with the pseudorapidity in the center-of-mass system.

The trigger conditions are slightly different for each of the three collision systems.
For \pp\ collisions, the VZERO detectors~\cite{Abelev:2013qoq} located in the forward and backward regions of the detector, 
as well as the Silicon Pixel Detector (SPD) at mid-rapidity are used to form a minimum-bias trigger by requiring 
at least one hit in the SPD or either of the VZERO detectors~\cite{Aamodt:2010pb}.  
For \PbPb\ and \pPb\ collisions, the trigger is formed by requiring simultaneous hits in both VZERO detectors.
In addition, high-multiplicity triggers in \pp\ and \pPb\ collisions based on the SPD are used.
Two additional triggers in \PbPb\ are used based on the VZERO signal amplitude which enhanced the statistics for central and semi-central 
collisions~\cite{Abelev:2013qoq}.
Approximately 164, 115, and 52 million events are used for \pp, \pPb, and \PbPb\ collisions, respectively.  
For pp and p--Pb, the high multiplicity triggers account for less than $3\%$ of the collected events.  
For Pb--Pb, the central and semi-central triggers account for about $40\%$ and $52\%$ of the collected events, respectively.

The Inner Tracking System (ITS) and Time Projection Chamber (TPC) located at mid-rapidity are used for particle tracking~\cite{Alme:2010ke}.
The ITS consists of 6 layers of silicon detectors: silicon pixel (layers 1,2), silicon drift (layers 3,4), 
and silicon strip (layers 5,6) detectors.
The ITS provides high spatial resolution of the primary vertex.
The TPC alone is used for momentum and charge determination of particles via their curvature in the 0.5~T longitudinal magnetic field, since cluster sharing within the ITS causes a small momentum bias for particle pairs at low relative momentum.

The TPC additionally provides particle identification capabilities through the specific ionization energy loss (\dEdx).  
The Time Of Flight (TOF) detector is also used to select particles at higher momenta.  
To ensure uniform tracking, the $z$-coordinate (beam-axis) of the primary vertex 
is required to be within a distance of $10$~cm from the detector center.
Events with less than three reconstructed charged pions are rejected, 
which removes about $25\%$ and $10\%$ of the low-multiplicity events in \pp\ and \pPb, respectively.

\section{Track selection and multiplicity intervals}
\label{sec:selection}
Tracks with total momentum less than 1.0 GeV/$c$ are used to ensure good particle identification.  
We also require transverse momentum $\pt>0.16$ GeV/$c$, and pseudorapidity \mbox{$|\eta|<0.8$}.  
To ensure good momentum resolution a minimum of 70 tracking points in the TPC are required.
Charged pions are selected if they are within 2 standard deviations~($\sigma$) of the expected pion \dEdx\ value~\cite{Abelev:2014ffa}.  
For momenta greater than 0.6 GeV/$c$, high purity is maintained with TOF by selecting particles within $2\sigma$ of the expected pion time-of-flight.
Additionally, tracks which are within $2\sigma$ of the expected kaon or proton \dEdx\ or time-of-flight values are rejected.
The effects of track merging and splitting are minimized based on the spatial separation of tracks in the TPC as described in~\cite{Abelev:2013pqa}.  
For three-pion correlations the pair cuts are applied to each of the three pairs in the triplet.

Similar as in \cite{Abelev:2013bla}, the analysis is performed in intervals of multiplicity which are defined by the reconstructed number of charged pions, $\Nrecp$, in the above-mentioned kinematic range. 
For each multiplicity interval, the corresponding mean acceptance and efficiency corrected value of the total charged-pion multiplicity, $\avNp$, and the total charged-particle multiplicity, $\Nch$, are determined using detector simulations with PYTHIA~\cite{Sjostrand:2006za}, DPMJET~\cite{Roesler:2000he}, 
and HIJING~\cite{Wang:1991hta} event generators.
The systematic uncertainty of $\Nch$ and $\avNp$ is determined by comparing PYTHIA to PHOJET~(\pp)~\cite{Engel:1994vs}, DPMJET to HIJING~(\pPb), and HIJING to AMPT~(\PbPb)~\cite{Lin:2004en}, and amounts to about $5\%$.
The multiplicity intervals, $\avNp$, $\Nch$, as well as the average centrality in \PbPb\ and fractional cross sections in \pp\ and \pPb\ are 
given in Table~\ref{tab:MultTable}.  The collision centrality in Pb--Pb is determined using the charged-particle multiplicity in the VZERO detectors~\cite{Abelev:2013qoq}.
As mentioned above, the center-of-mass reference frame for \pPb\ collisions does not coincide with the laboratory frame, where $\Nch$ is measured.
However, from studies using DPMJET and HIJING at the generator level, 
the difference to $\Nch$ measured in the center-of-mass is expected to be smaller than $3\%$.

\begin{table}
  \footnotesize
  \center
  \begin{tabular}{ l | c c c | c c c | c c c } 
    & \multicolumn{3}{c|}{\PbPb\ data} & \multicolumn{3}{c|}{\pPb\ data} & \multicolumn{3}{c}{\pp\ data} \\ 
    $\Nrecp$ & $\left<\rm{Cent}\right>$ & $\avNp$ & $\Nch$ & Fraction & $\avNp$ & $\Nch$ & Fraction & $\avNp$ & $\Nch$  \\ \hline 
$[3,5)$       & -     & - & -          & 0.10 & - & -             & 0.23 & 4.0 & 4.6   \\ 
$[5,10)$      & -     & - & -          & 0.20  & 8.5 & 9.8      & 0.31 & 7.7 & 8.6   \\
$[10,15)$     & -     & - & -          & 0.18  & 15 & 17       & 0.12 & 13 & 15   \\
$[15,20)$     & -     & - & -          & 0.14  & 20 & 23       & 0.05 & 18 & 20   \\
$[20,30)$     & 77\%  & 26 & 36        & 0.17  & 29 & 33       & 0.03 & 24 & 27   \\
$[30,40)$     & 73\%  & 37 & 50        & 0.07  & 40 & 45       & 0.003 & 34 & 37  \\
$[40,50)$     & 70\%  & 49 & 64        & 0.03  & 51 & 57       & $1\times10^{-4}$ & 44 & 47        \\ 
$[50,70)$     & 66\%  & 66 & 84        & 0.01 & 63 & 71       & - & - & -        \\ 
$[70,100)$    & 60\%  & 95 & 118       & - & - & -             & - & - & -        \\ 
$[100,150)$   & 53\%  & 142 & 172      & - & - & -             & - & - & -        \\ 
$[150,200)$   & 48\%  & 213 & 253      & - & - & -             & - & - & -        \\ 
$[200,260)$   & 43\%  & 276 & 326      & - & - & -             & - & - & -        \\ 
$[260,320)$   & 38\%  & 343 & 403      & - & - & -             & - & - & -        \\ 
$[320,400)$   & 33\%  & 426 & 498      & - & - & -             & - & - & -        \\ 
$[400,500)$   & 28\%  & 534 & 622      & - & - & -             & - & - & -        \\ 
$[500,600)$   & 22\%  & 654 & 760      & - & - & -             & - & - & -        \\ 
$[600,700)$   & 18\%  & 777 & 901      & - & - & -             & - & - & -        \\ 
$[700,850)$   & 13\%  & 931 & 1076     & - & - & -             & - & - & -        \\ 
$[850,1050)$  & 7.4\% & 1225 & 1413     & - & - & -             & - & - & -        \\ 
$[1050,2000)$ & 2.6\% & 1590 & 1830    & - & - & -             & - & - & -        \\ 
  \end{tabular}
  \caption{Multiplicity intervals as determined by the reconstructed number of charged pions, $\Nrecp$, with all of the track selection 
           cuts~($p<1.0$ GeV/$c$, $\pt>0.16$ GeV/$c$, $|\eta|<0.8$).
           $\avNp$ stands for the acceptance corrected average number of charged pions, and $\Nch$
           for corresponding acceptance corrected number of charged particles in the same kinematic range.
           The uncertainties on $\Nch$ are about $5\%$.  
           The RMS width of the $\Nch$ distribution in each interval ranges from $10\%$ to $35\%$ for the highest and lowest multiplicity intervals, respectively.
           The average centrality for \PbPb\ in percentiles, as well as the fractional cross-sections of the multiplicity intervals for \pPb\ and \pp\ 
           are also given.
           The RMS widths for the centralities range from about $2$ to $4$ percentiles for central and peripheral collisions, respectively.} 
  \label{tab:MultTable}
\end{table}

\section{Analysis technique}
\label{sec:analysis}
To extract the source radii, one can measure two- and three-particle correlation functions as in Ref~\cite{Abelev:2013pqa}.
The two-particle correlation function 
\begin{equation}
  C_2(p_1,p_2) = \alpha_2 \, \frac{N_2(p_1,p_2)}{N_1(p_1)\,N_1(p_2)} 
  \label{eq:C2}
\end{equation}
is constructed using the momenta $p_i$, and is defined as the ratio of the inclusive two-particle spectrum over the product of the inclusive single-particle spectra.
Both are projected \com{separately} onto the Lorentz invariant relative momentum $\qinv = \sqrt{-(p_1-p_2)^{\mu}(p_1-p_2)_{\mu}}$ 
and the average pion transverse momentum $\kT=|\vec{p}_{\rm T,1}+\vec{p}_{\rm T,2}|$/2.
The numerator of the correlation function is formed by all pairs of particles from the same event.  
The denominator is formed by taking one particle from one event and the second particle from another event within the same multiplicity interval.  
The normalization factor, $\alpha_2$, is determined such that the correlation function equals unity in a certain interval of relative momentum $q$.  The location of the interval is sufficiently above the dominant region 
of QS+FSI correlations and sufficiently narrow to avoid the influence of non-femtoscopic correlations at large relative momentum.  
As the width of QS+FSI correlations is different in all three collision systems, our choice for the normalization interval depends on the multiplicity interval.
For \PbPb, the normalization intervals are $0.15<q<0.175$ GeV/$c$ for $\Nrecp\geq400$ and $0.3<\qinv<0.35$ GeV/$c$ for $\Nrecp<400$.  
For \pp\ and \pPb\ the normalization interval is $1.0<\qinv<1.2$ GeV/$c$.

Following~\cite{Bowler:1991vx,Sinyukov:1998fc}, the two-particle QS distributions, $N_2^{\rm QS}$, and correlations, $C_2^{\rm QS}$, are extracted 
from the measured distributions in intervals of \kT~assuming
\begin{equation}
\centering
C_2(\qinv) = {\mathcal N}\,[(1-\fc^2) + \fc^2\,K_2(\qinv)\,C_2^{\rm QS}(\qinv)]\,B(\qinv)\,.
\label{eq:C2QS}
\end{equation}
The parameter $\fc^2$ characterizes the combined dilution effect of weak decays and long-lived resonance decays in the ``core/halo'' picture~\cite{Lednicky:1979ig,Csorgo:1994in}. 
In \PbPb, it was estimated to be $0.7\pm0.05$ with mixed-charge two-pion correlations~\cite{Abelev:2013pqa}.
The same procedure performed in pp and p--Pb data results in compatible values.
The FSI correlation is given by $K_2(\qinv)$, which includes Coulomb and strong interactions.
For low multiplicities ($\Nrecp<150$), $K_2(\qinv)$ is calculated iteratively using the Fourier transform of the FSI corrected correlation functions.
For higher multiplicities~($\Nrecp\geq150$), $K_2(\qinv)$ is calculated as in Ref.~\cite{Abelev:2013pqa} 
using the \mbox{THERMINATOR2} model~\cite{Kisiel:2005hn, Chojnacki:2011hb}. 
$B(\qinv)$ represents the non-femtoscopic background correlation, and is taken from PYTHIA and DPMJET 
for \pp\ and \pPb, respectively~\cite{Aamodt:2010jj,Aamodt:2011kd}. 
It is set equal to unity for \PbPb, where no significant background is expected.
In Eq.~\ref{eq:C2QS}, ${\mathcal N}$ is the residual normalization of the fit which typically differs from unity by 0.01.

The same-charge two-pion QS correlation can be parametrized by an exponential
\begin{equation}
\centering
C_2^{\rm QS}(\qinv) =  1 + \lambda\,e^{-\Rinv\,\qinv}\,, \label{eq:C2ssExp}
\end{equation}
as well as by a Gaussian or Edgeworth expansion
\begin{eqnarray}
\centering
C_2^{\rm QS}(\qinv) &=& 1 + \lambda\,E_{\rm w}^2(\Rinv\,\qinv)\,e^{-\Rinv^2\,\qinv^2} \label{eq:C2ssparameters} \\
E_{\rm w}(\Rinv\,\qinv) &=& 1 + \sum_{n=3}^{\infty} \frac{\kappa_n}{n! (\sqrt{2})^n} H_n(\Rinv\,\qinv)\,,
\end{eqnarray}
where $E_{\rm w}(\Rinv\qinv)$ characterizes deviations from Gaussian behavior, 
$H_n$ are the Hermite polynomials, 
and $\kappa_n$ are the Edgeworth coefficients~\cite{Csorgo:2000pf}.  
The first two relevant Edgeworth coefficients~($\kappa_3, \kappa_4$) are found to be sufficient to describe the non-Gaussian features at low relative momentum.
The Gaussian functional form is obtained with $E_{\rm w}=1$~($\kappa_n=0$) in Eq.~\ref{eq:C2ssparameters}.
The parameter $\lambda$ characterizes an apparent suppression from an incorrectly assumed functional form of $C_2^{\rm QS}$ and the suppression due to possible pion coherence~\cite{Akkelin:2001nd}.
The parameter $\Rinv$ is the characteristic radius from two-particle correlations evaluated in the pair-rest frame.
The effective intercept parameter for the Edgeworth fit is given by $\ls=\lambda\,E_{\rm w}^2(0)$~\cite{Csorgo:2000pf}.
The effective intercept can be below the chaotic limit of 1.0 for partially coherent emission~\cite{Andreev:1992pu,Akkelin:2001nd,Abelev:2013pqa}.
The extracted effective intercept parameter is found to strongly depend on the assumed functional form of $C_2^{\rm QS}$.

The three-particle correlation function 
\begin{equation}
\centering
  C_3(p_1,p_2,p_3) = \alpha_3 \, \frac{N_3(p_1,p_2,p_3)}{N_1(p_1)\,N_1(p_2)\,N_1(p_3)} \label{eq:C3}
\end{equation}
is defined as the ratio of the inclusive three-particle spectrum over the product 
of the inclusive single-particle spectra. 
In analogy to the two-pion case, it is projected onto the Lorentz invariant $Q_3 = \sqrt{q_{12}^2+q_{31}^2+q_{23}^2}$
and the average pion transverse momentum $\KTThree = \frac{|\vec{p}_{\rm T,1}+\vec{p}_{\rm T,2}+\vec{p}_{\rm T,3}|}{3}$.
The numerator of $C_3$ is formed by taking three particles from the same event. 
The denominator is formed by taking each of the three particles from different events. 
The normalization factor, $\alpha_3$, is determined such that the correlation function equals unity in the interval of $Q_3$ where each pair $\qinv_{ij}$ lies in the same interval given before for two-pion correlations.

The extraction of the full three-pion QS distribution, $N_3^{\rm QS}$, in intervals of $\KTThree$ is done as in Ref~\cite{Abelev:2013pqa}
by measuring
\begin{eqnarray}
\centering
N_3(p_1,p_2,p_3) &=& f_1\,N_1(p_1)\,N_1(p_2)\,N_1(p_3) \nonumber \\
&+& f_2\,\left[N_2(p_1,p_2)\,N_1(p_3) + N_2(p_3,p_1)\,N_1(p_2) + N_2(p_2,p_3)\,N_1(p_1) \right] \nonumber \\
&+& f_3\,K_3(q_{12}, q_{31}, q_{23})\,N_3^{\rm QS}(p_1,p_2,p_3)\,,
\label{eq:N3QS}
\end{eqnarray}
where the fractions $f_1=(1-f_c)^3 + 3f_c(1-f_c)^2 - 3(1-f_c)(1-f_c^2)=-0.08$, $f_2=1-f_c=0.16$, and $f_3=f_c^3=0.59$ using $f_c^2=0.7$ as in the two-pion case.
The term $N_2(p_i,p_j)\,N_1(p_k)$ is formed by taking two particles from the same event and the third particle from a mixed event.  All three-particle distributions are normalized to each other in the same way as for $\alpha_3$.
$K_3(q_{12}, q_{31}, q_{23})$ denotes the three-pion FSI correlation, which in the generalized Riverside~(GRS) approach~\cite{Aggarwal:2000ex,Adams:2003vd,Abelev:2013pqa}
is approximated by $K_2(q_{12})\,K_2(q_{31})\,K_2(q_{23})$.  It was found to describe the $\pi^{\pm}\pi^{\pm}\pi^{\mp}$ 
three-body FSI correlation to the few percent level~\cite{Abelev:2013pqa}.
From Eq.~\ref{eq:N3QS} one can extract $N_3^{\rm QS}$ and construct the three-pion QS cumulant correlation 
\begin{eqnarray}
\centering
{\rm {\bf c}}_3(p_1,p_2,p_3) &=& {\mathcal N}_3\,[1 + [2N_1(p_1)\,N_1(p_2)\,N_1(p_3) \nonumber \\
&-& N_2^{\rm QS}(p_1,p_2)\,N_1(p_3)-N_2^{\rm QS}(p_3,p_1)\,N_1(p_2)-N_2^{\rm QS}(p_2,p_3)\,N_1(p_1) \nonumber \\
&+& N_3^{\rm QS}(p_1,p_2,p_3)]/N_1(p_1)\,N_1(p_2)\,N_1(p_3)\,],
\label{eq:c3}
\end{eqnarray}
where $N_2^{\rm QS}(p_i,p_j)\,N_1(p_k) = [N_2(p_i,p_j)\,N_1(p_k) - N_1(p_i)N_1(p_j)N_1(p_k)(1-f_c^2)]/(f_c^2K_2)$.
In Eq.~\ref{eq:c3}, all two-pion QS correlations are explicitly subtracted~\cite{Heinz:2004pv}. 
The QS cumulant in this form has FSIs removed before its construction.
${\mathcal N}_3$ is the residual normalization of the fit which typically differs from unity by 0.02.

The three-pion same-charge cumulant correlations are then projected onto 3D pair relative momenta 
and fit with an exponential
\begin{equation}
{\rm {\bf c}}_3(\qinv_{12},\qinv_{31},\qinv_{23}) = 1 + \lambda_3 \,e^{-\Rinvt\,(\qinv_{12} + \qinv_{31} + \qinv_{23})/2}\,,
\end{equation}
as well as a Gaussian and an Edgeworth expansion~\cite{Csorgo:2000pf}
\begin{eqnarray}
{\rm {\bf c}}_3(\qinv_{12},\qinv_{31},\qinv_{23}) &=& 1 + \lambda_3 \,  
                                                    E_{\rm w}(\Rinvt\,\qinv_{12})\,E_{\rm w}(\Rinvt\,\qinv_{31})\,E_{\rm w}(\Rinvt\,\qinv_{23}) \, e^{-\Rinvt^2\,Q_3^2/2}.
\label{eq:c3Fit2}
\end{eqnarray}
$\Rinvt$ and $\lambda_3$ are the invariant radius and intercept parameters extracted from three-pion cumulant correlations, respectively.  The effective intercept parameter for the Edgeworth fit is $\lst=\lambda_3\,E_{\rm w}^3(0)$.  For an exact functional form of ${\rm {\bf c}}_3$, $\lst$ reaches a maximum of 2.0 for fully chaotic pion emission.  
Deviations below and above 2.0 can further be caused by incorrect representations of ${\rm {\bf c}}_3$, e.g.\ Gaussian.
Equation \ref{eq:c3Fit2} neglects the effect of the three-pion phase~\cite{Heinz:1997mr} which was found to be consistent with zero for \PbPb\ central and mid-central collisions \cite{Abelev:2013pqa}.  
We note that the extracted radii from two-and three-pion correlations need not exactly agree, e.g.\ in the case of coherent emission~\cite{Plumer:1992au}.

\begin{figure}[t]
  \centering
  \includegraphics[width=0.5\textwidth]{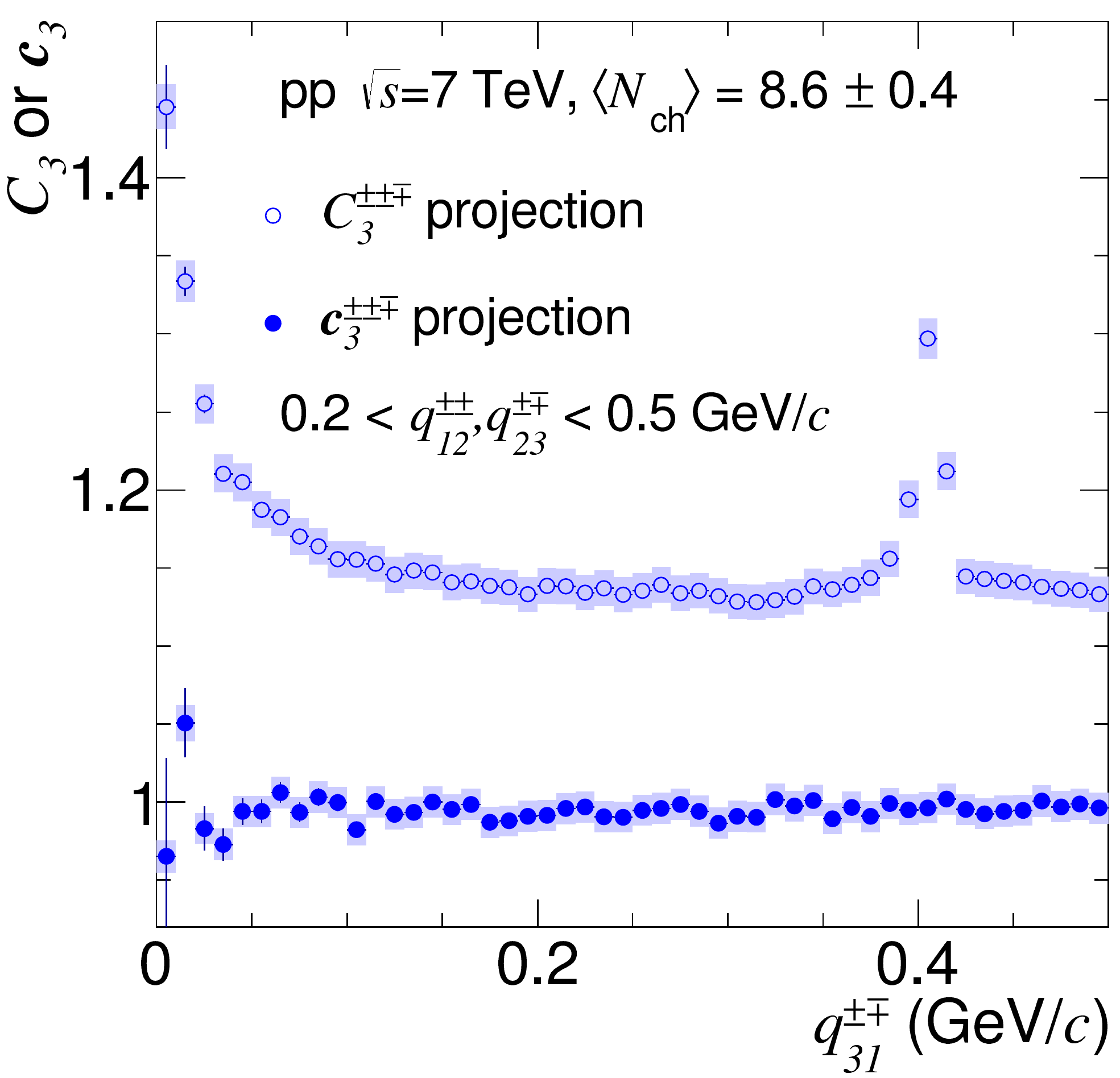}
  \caption{Demonstration of the removal of the $K^0_s$ decay from three-pion cumulants.  
           Mixed-charge three-pion correlations are projected against the relative momentum of a mixed-charge pair ($q_{31}^{\pm\mp}$).  
           The $K^0_s$ decay into a $\pi^{+}+\pi^{-}$ pair is visible as expected around 0.4 GeV/$c$.
           The FSI enhancement of the mixed-charge pair ``31" is also visible at low $q_{31}^{\pm\mp}$.
           FSI corrections are not applied.  
           Systematic uncertainties are shown by shaded boxes.}
  \label{fig:K0sRemoval}
\end{figure}

The measured correlation functions \com{$C_2$ and $C_3$}need to be corrected for finite track momentum resolution of the TPC which causes a slight broadening of the correlation functions and leads to a slight decrease of the extracted radii.
PYTHIA~(\pp), DPMJET~(\pPb) and HIJING~(\PbPb) simulations are used to estimate the effect
on the fit parameters.  After the correction, both fit parameters increase by about $2$\%~($5$\%) for the lowest~(highest) multiplicity interval.
The relative systematic uncertainty of this correction is conservatively taken to be $1$\%. 
The pion purity is estimated to be about $96$\%. 
Muons are found to be the dominant source of contamination, for which we apply corrections to the correlation functions as described in \Ref{Abelev:2013pqa}.
The correction typically increases the radius~(intercept) fit parameters by less than $1\%$~($5$\%).
The corresponding systematic uncertainty is included in the comparison of the mixed-charged correlation with unity~(see below).

\begin{figure}[t]
  \centering
  \includegraphics[width=0.95\textwidth]{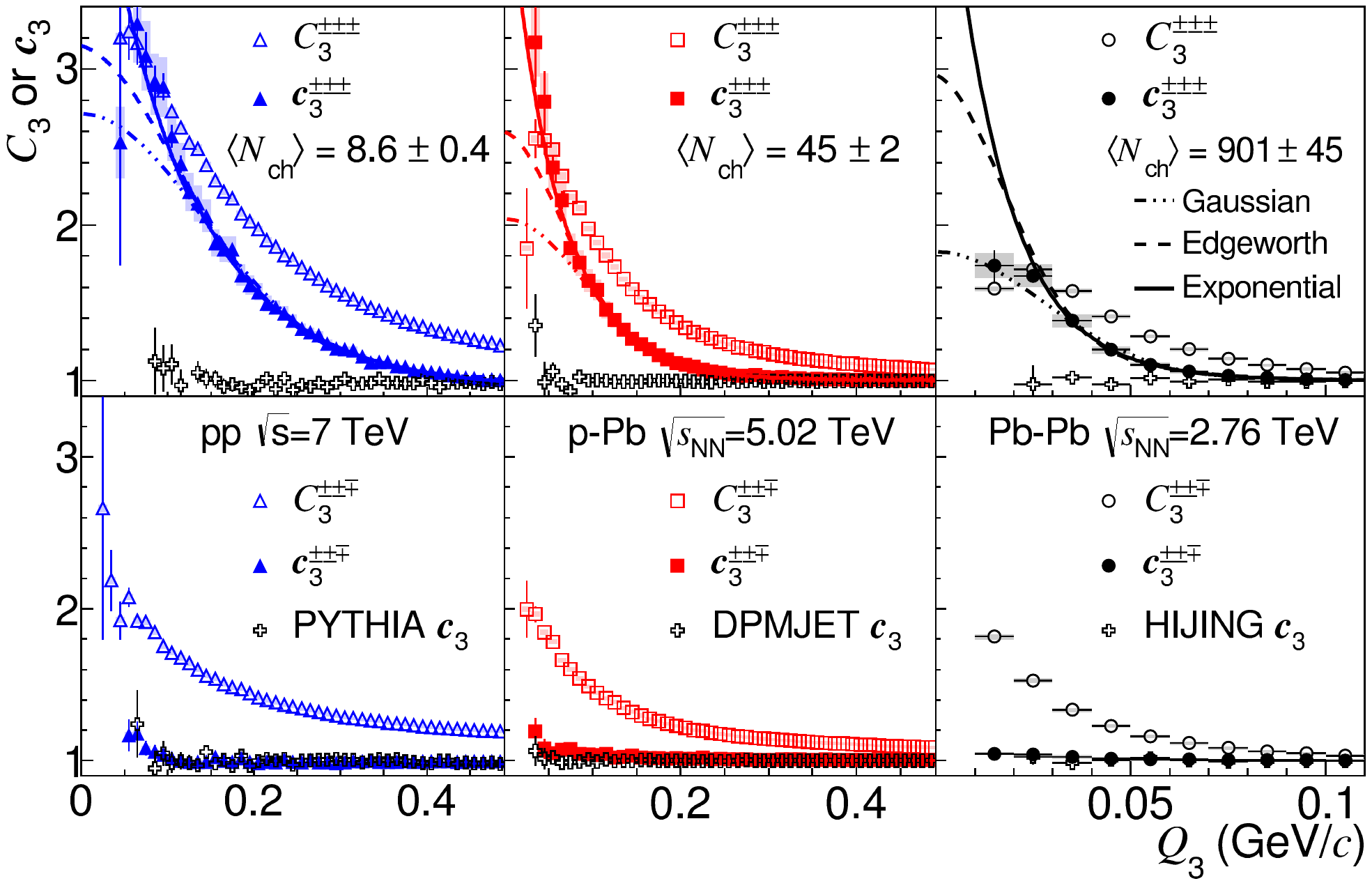}
  \caption{Three-pion correlation functions versus $Q_3$ for $0.16<\KTThree<0.3$ GeV/$c$ in \pp, \pPb\ and \PbPb\ 
           collision data compared to PYTHIA, DPMJET and HIJING generator-level calculations. 
           Top panels are for same-charge triplets, while bottom panels are for mixed-charge triplets. Two points at low $Q_3$ 
           with large statistical uncertainties are not shown for the \pp~same-charge correlation function.
           }
  \label{fig:CorrelationFunctions}
\end{figure}

\section{Results}
\label{sec:results}
The absence of two-particle correlations in the three-pion cumulant can be demonstrated via the removal of known two-body effects such as the decay of $K^0_s$ into a $\pi^+ + \pi^-$ pair~(\Fig{fig:K0sRemoval}).
The mixed-charge three-pion correlation function ($C_3^{\pm\pm\mp}$) projected onto the invariant relative momentum 
of one of the mixed-charge pairs in the triplet exhibits the $K^0_s$ peak as expected around $q^{\pm\mp}=0.4$ GeV/$c$, while it is removed in the cumulant.

In \Fig{fig:CorrelationFunctions} we present three-pion correlation functions for same-charge~(top panels) and mixed-charge~(bottom panels)
triplets in \pp, \pPb, and \PbPb\ collision systems in three sample multiplicity intervals.  
For same-charge triplets, the three-pion cumulant QS correlation~(${\rm {\bf c}}_3^{\pm\pm\pm}$) is clearly visible. 
For mixed-charge triplets the three-pion 
cumulant correlation function~(${\rm {\bf c}}_3^{\pm\pm\mp}$) is consistent with unity, as expected when FSIs are removed.
Gaussian, Edgeworth, and exponential fits are performed in three dimensions~($q_{12},q_{31},q_{23}$). 
Concerning Edgeworth fits, different values of the $\kappa$ coefficients correspond to different spatial freeze-out profiles.  
In order to make a meaningful comparison of the characteristic radii across all multiplicity intervals and collision systems, 
we fix $\kappa_3=0.1$ and $\kappa_4=0.5$.
The values are determined from the average of free fits to ${\rm {\bf c}}_3^{\pm\pm\pm}$ for all multiplicity intervals, 
\KTThree\ intervals and systems.  The RMS of both $\kappa_3$ and $\kappa_4$ distributions is 0.1.
The chosen $\kappa$ coefficients produce a sharper correlation function which corresponds to larger tails in the source distribution.
Also shown in Fig.~\ref{fig:CorrelationFunctions} are model calculations of ${\rm {\bf c}}_3$ in PYTHIA (\pp), DPMJET (\pPb) and HIJING (\PbPb),
which do not contain QS+FSI correlations and demonstrate that three-pion cumulants, in contrast to two-pion correlations~\cite{Aamodt:2010jj,Aamodt:2011kd}, do not contain a significant non-femtoscopic background, even for low multiplicities. 

The systematic uncertainties on $C_3$\com{in Fig.~\ref{fig:CorrelationFunctions}} are conservatively estimated to be $1\%$ by comparing 
$\pi^+$ to $\pi^-$ correlation functions and by tightening the track merging and splitting cuts. 
The systematic uncertainty on ${\rm {\bf c}}_3^{\pm\pm\pm}$ is estimated by the residual correlation observed with ${\rm {\bf c}}_3^{\pm\pm\mp}$ relative to unity.  
The residual correlation leads to a $4\%$ uncertainty on $\lst$ while having a negligible effect on $\Rinvt$.
The uncertainty on $\fc$ leads to an additional $10\%$ uncertainty on ${\rm {\bf c}}_3-1$ and $\lst$.
We also investigated the effect of setting $f_c=1$ and thus $f_1=0, f_2=0, f_3=1.0$ in Eq.~\ref{eq:N3QS} and found a negligible effect on $\Rinvt$, while significantly reducing $\lst$ as expected when the dilution is not adequately taken into account.

\begin{figure}[t]
  \centering
  \subfigure[Low \kT~and \KTThree]{
    \includegraphics[width=0.47\textwidth]{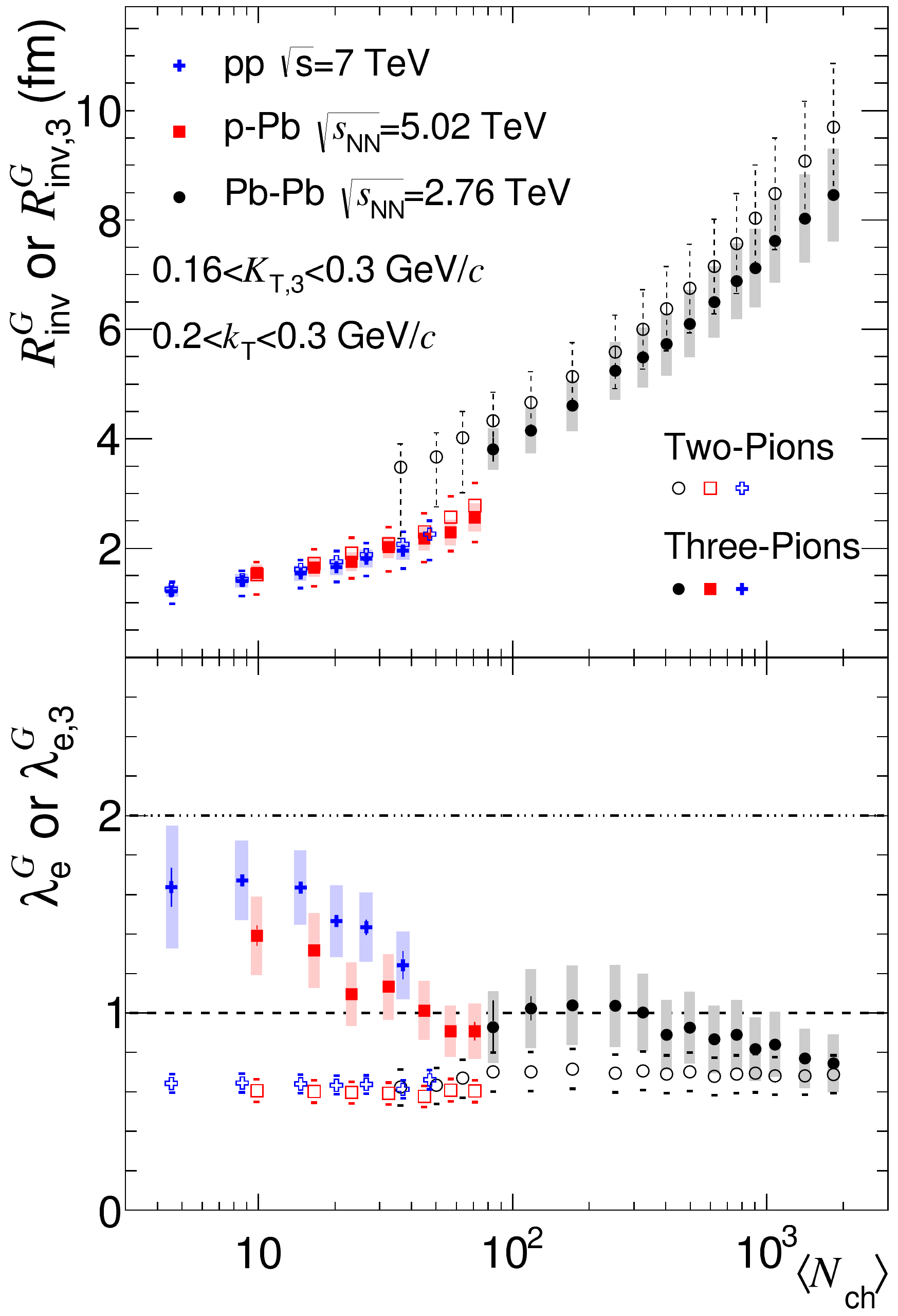}
    \label{fig:ParamGaussK1}
  } \hspace{0.1cm}
  \subfigure[High \kT~and \KTThree]{
    \includegraphics[width=0.47\textwidth]{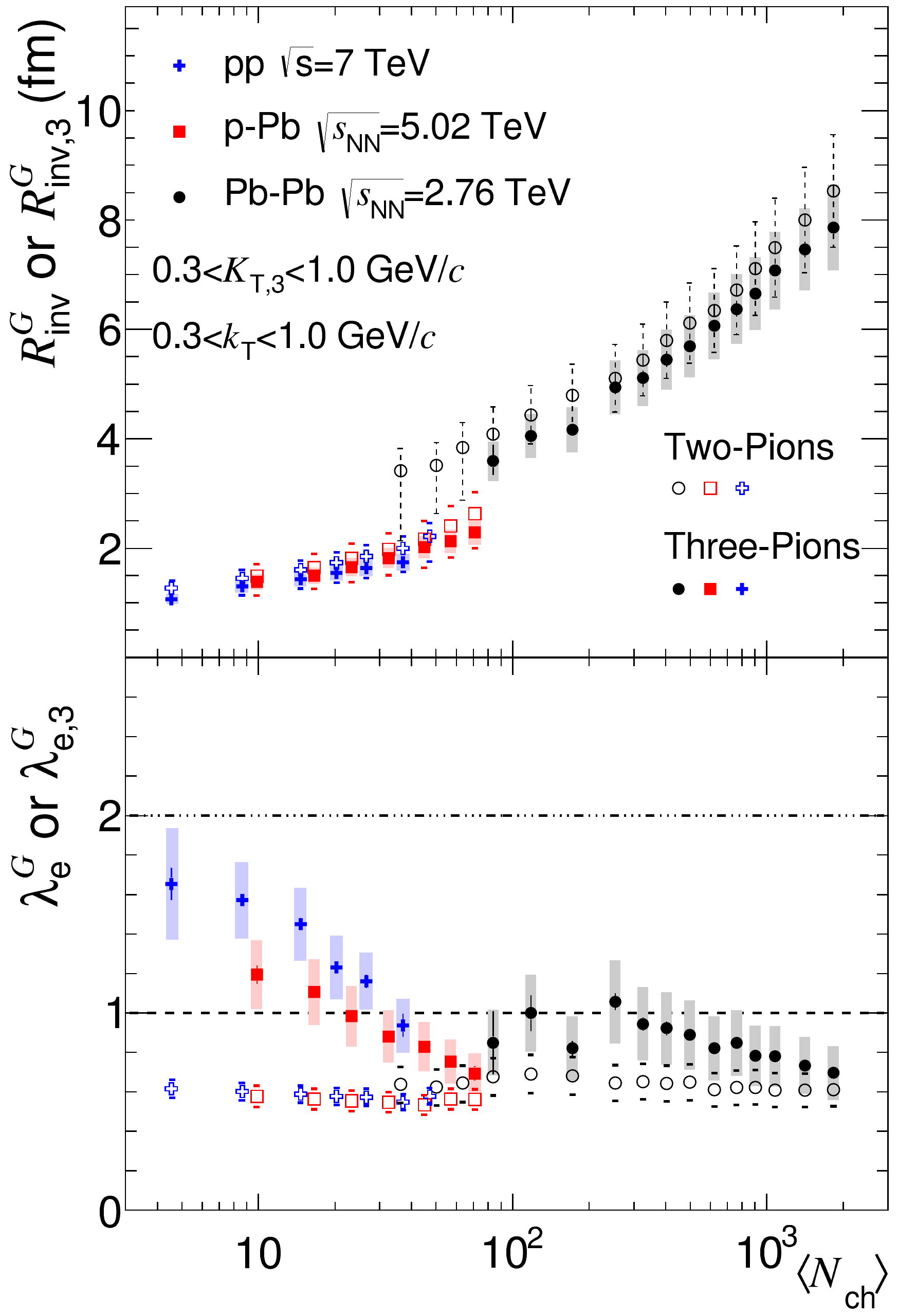}
    \label{fig:ParamGaussK2}
  }
  \caption{Two- and three-pion Gaussian fit parameters versus $\Nch$ in \pp, \pPb\ and \PbPb\ collision systems 
           for low and high \kT\ and \KTThree~intervals.  
           Top panels show the Gaussian radii $R^{\rm G}_{\rm inv}$ and $R^{\rm G}_{\rm inv,3}$ 
           and bottom panels show the effective Gaussian intercept parameters $\lambda^{\rm G}_{\rm e}$ and $\lambda^{\rm G}_{\rm e,3}$.
           The systematic uncertainties are dominated by fit range variations and are shown by bounding/dashed lines and shaded boxes for two- and three-particle parameters, respectively.  
           The dashed and dash-dotted lines represent the chaotic limits for $\lambda^{\rm G}_{\rm e}$ and $\lambda^{\rm G}_{\rm e,3}$, respectively.}
  \label{fig:ParamGauss}
\end{figure}

Figures~\ref{fig:ParamGaussK1} and \ref{fig:ParamGaussK2} show the three-pion Gaussian fit parameters for low and high $K_{T,3}$ intervals, respectively.
The $\left<k_{\rm T}\right>$ values for low~(high) $\kT$ are 0.25~(0.43) GeV/$c$.
The resulting pair \kT\ distributions in the triplet \KTThree~intervals have RMS widths for the low~(high) $\KTThree$ of $0.12$~($0.14$) in \pp\ and \pPb\ and $0.04$~($0.09$) GeV/$c$ in \PbPb\ collisions. The $\left<k_{\rm T}\right>$ values for low~(high) $\KTThree$ are 0.24~(0.39) GeV/$c$. 
We also show the fit parameters extracted from two-pion correlations in order to compare to those extracted from three-pion cumulants. 
For \PbPb, the Gaussian radii extracted from three-pion correlations are about $10\%$ smaller than those extracted from two-pion correlations, which
may be due to the non-Gaussian features of the correlation function. 
A clear suppression below the chaotic limit is observed for 
the effective intercept parameters in all multiplicity intervals.
The suppression may be caused by non-Gaussian features of the correlation function and also by a finite coherent component of pion emission~\cite{Andreev:1992pu,Akkelin:2001nd,Abelev:2013pqa}.

\begin{figure}[t]
  \centering
  \subfigure[Low \kT~and \KTThree]{
    \includegraphics[width=0.47\textwidth]{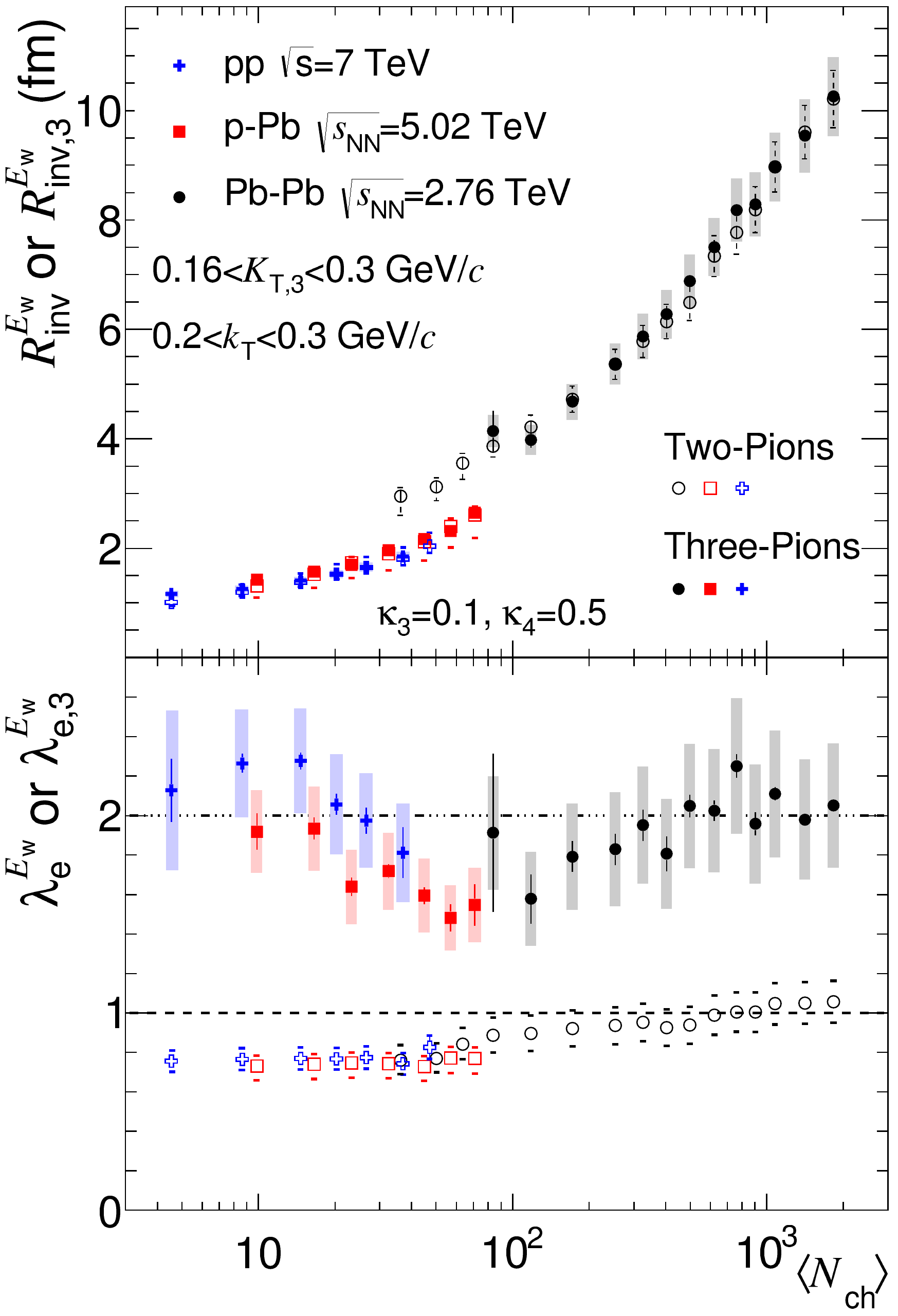}
    \label{fig:ParamEWK1}
  } \hspace{0.1cm}
  \subfigure[High \kT~and \KTThree]{
    \includegraphics[width=0.47\textwidth]{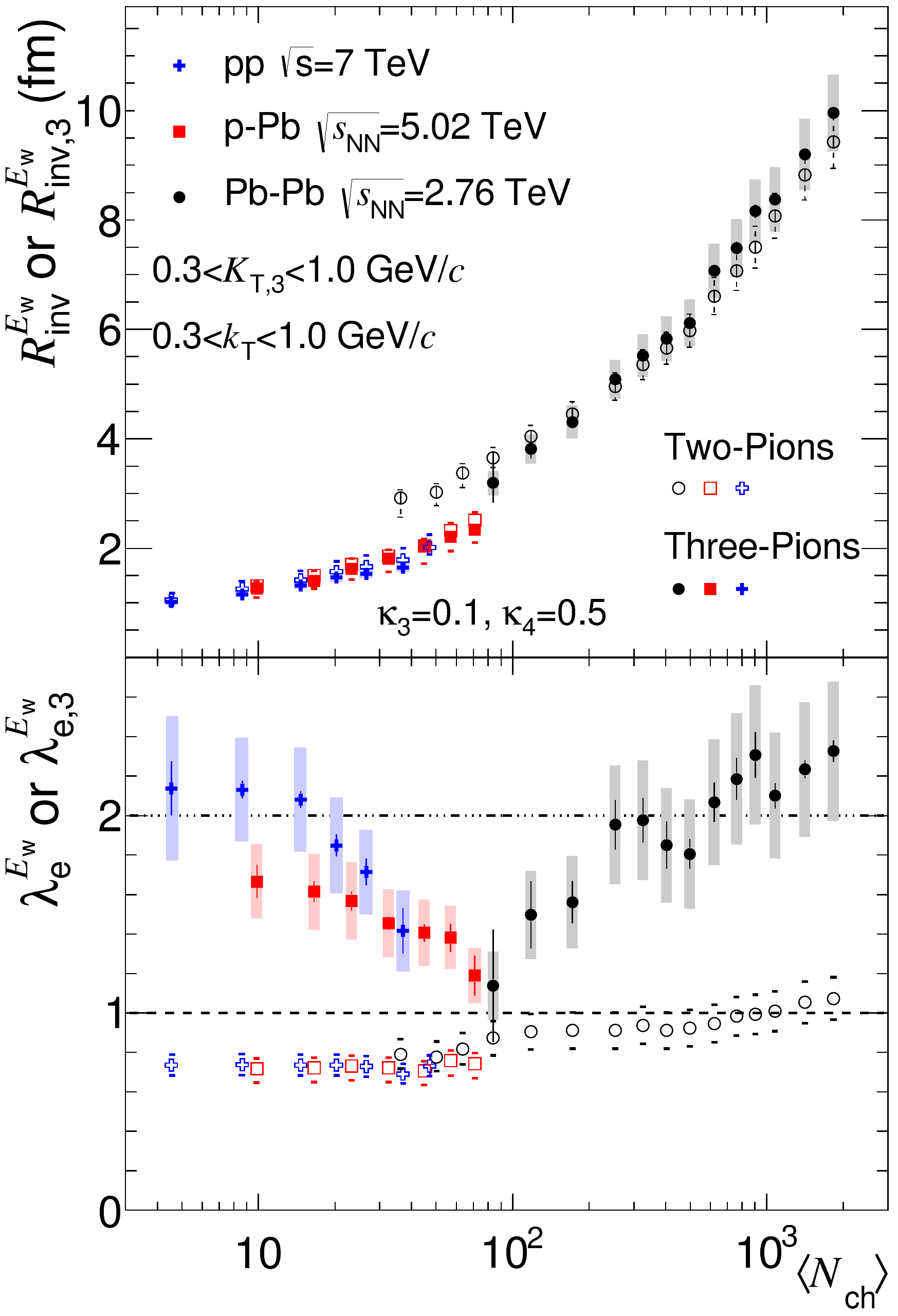}
    \label{fig:ParamEWK2}
  }
  \caption{Two- and three-pion Edgeworth fit parameters versus $\Nch$ in \pp, \pPb\ and \PbPb\ collision systems 
           for low and high \kT~and \KTThree~intervals.  
           Top panels show the Edgeworth radii $R^{E_{\rm w}}_{\rm inv}$ and $R^{E_{\rm w}}_{\rm inv,3}$ and 
           bottom panels show the effective intercept parameters $\lambda^{E_{\rm w}}_{\rm e}$ and $\lambda^{E_{\rm w}}_{\rm e,3}$.  
           As described in the text, $\kappa_3$ and $\kappa_4$ are fixed to 0.1 and 0.5, respectively.
           Same details as for Fig.~\ref{fig:ParamGauss}.
  } \label{fig:ParamEW}
\end{figure}

The systematic uncertainties on the fit parameters are dominated by fit-range variations, 
especially in the case of Gaussian fits to non-Gaussian correlation functions.  
The chosen fit range for ${\rm {\bf c}}_3$ varies smoothly between $Q_3=0.5$ and $0.1$ GeV/$c$ 
from the lowest multiplicity \pp\ to the highest multiplicity \PbPb\ intervals. 
For $C_2$, the fit ranges are chosen to be $\sqrt{2}$ times narrower.  The characteristic width of Gaussian three-pion cumulant QS correlations projected against $Q_3$ is a factor of $\sqrt{2}$ times that of Gaussian two-pion QS correlations projected against $\qinv$~\cite{Weiner:1988jy,Andreev:1992pu}.
As a variation we change the upper bound of the fit range by $\pm30\%$ for three-pion correlations and two-pion correlations in \PbPb~for $\Nrecp>50$.  
For $\Nrecp<50$, in Pb--Pb, the upper limit of the fit range is increased to match that in p--Pb (i.e.\ 0.13 to 0.27 GeV/$c$). 
For \pp\ and \pPb, owing to the larger background present for two-pion correlations, we extend the fit range to $\qinv=1.2$ GeV/$c$ for the upper variation. 
The non-femtoscopic background in Eq.~\ref{eq:C2QS} has a non-negligible effect on the extracted radii in the extended fit range.
The resulting systematic uncertainties are fully correlated for three-pion fit parameters for each collision system, since the fit-range variations have the same 
effect in each multiplicity interval.  
The systematic uncertainties for the two-pion fit parameters are largely correlated and are asymmetric due to the different fit-range variations.
We note that the radii in \pp\ collisions at $\sqrt{s}=7$ TeV from our previous two-pion measurement~\cite{Aamodt:2011kd} are about $25\%$ smaller 
than the central values extracted in this analysis although compatible within systematic uncertainties. 
The large difference is attributed to the narrower fit range in this analysis.
In~\cite{Aamodt:2010jj,Aamodt:2011kd} the chosen Gaussian fit range was $\qinv<1.4$ GeV/$c$, while here it is $\qinv<0.35$ GeV/$c$ for the lowest multiplicity interval.  
The narrower fit range is chosen based on observations made with three-pion cumulants for which two-pion background correlations are removed.  
It is observed in Fig.~\ref{fig:CorrelationFunctions} that even for low multiplicities, the dominant QS correlation is well below $Q_3=0.5$ GeV/$c$.
The presence of the non-femtoscopic backgrounds can also bias the radii from two-pion correlations in wide fit ranges 
and is suppressed with three-pion cumulant correlations.

\begin{figure}[t]
  \centering
  \subfigure[Low \kT~and \KTThree]{
    \includegraphics[width=0.47\textwidth]{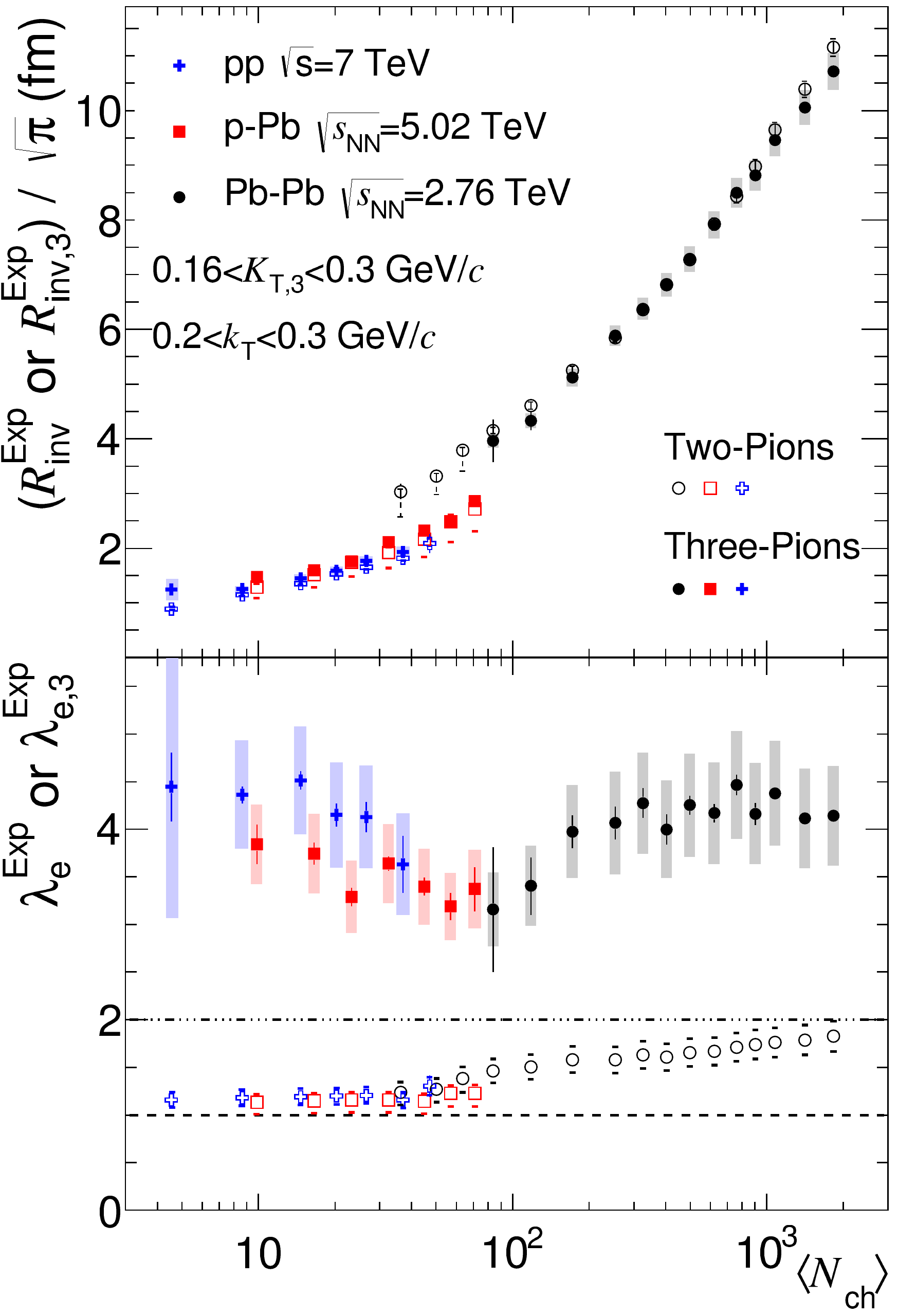}
    \label{fig:ParamExpK1}
  } \hspace{0.1cm}
  \subfigure[High \kT~and \KTThree]{
    \includegraphics[width=0.47\textwidth]{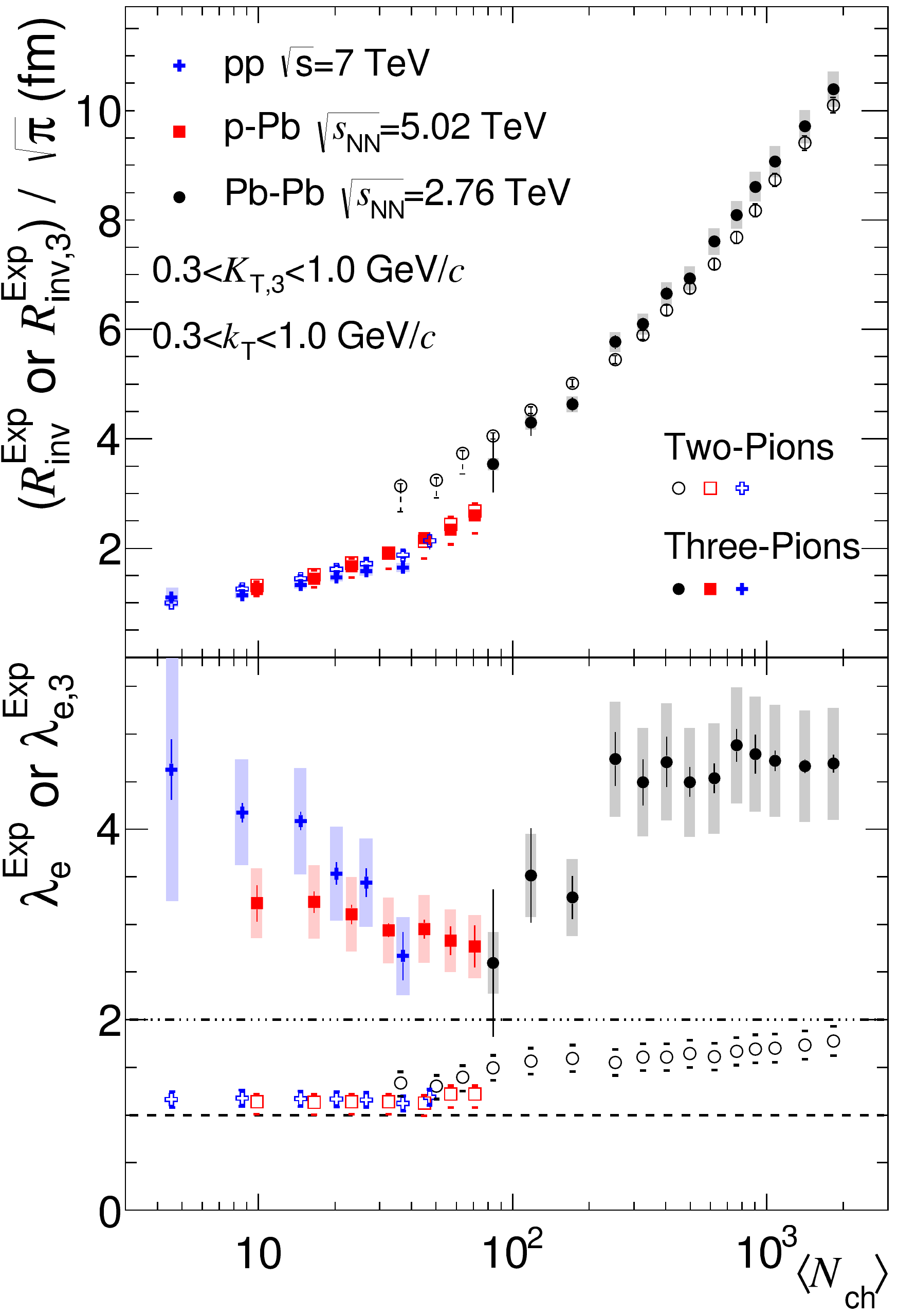}
    \label{fig:ParamExpK2}
  }
  \caption{Two- and three-pion Exponential fit parameters versus $\Nch$ in \pp, \pPb\ and \PbPb\ collision systems 
           for low and high \kT~and \KTThree~intervals.  
           Top panels show the Exponential radii $R^{\rm Exp}_{\rm inv}$ and $R^{\rm Exp}_{\rm inv,3}$ scaled down by $\sqrt{\pi}$ and 
           bottom panels show the effective intercept parameters $\lambda^{\rm Exp}_{\rm e}$ and $\lambda^{\rm Exp}_{\rm e,3}$.  
           Same details as for Fig.~\ref{fig:ParamGauss}.
  } \label{fig:ParamExp}
\end{figure}
To further address the non-Gaussian features of the correlation functions, we also extract the fit parameters from an Edgeworth and exponential parametrization as shown in Figs.~\ref{fig:ParamEW} and \ref{fig:ParamExp}.
We observe that the Edgeworth and exponential radii are significantly larger than the Gaussian radii.
However, they should not be directly compared as they correspond to different source profiles.
Gaussian radii correspond to the standard deviation of a Gaussian source profile whereas exponential radii correspond to the FWHM of a Cauchy source.  
The Edgeworth radii are model independent and are defined as the $2^{nd}$ cumulant of the measured correlation function.
Note that the exponential radii have been scaled down by $\sqrt{\pi}$ as is often done to compare Gaussian and exponential radii~\cite{Khachatryan:2010un}.
Compared to the Gaussian radii, the two- and three-pion radii are in much better agreement for the Edgeworth and exponential fits. 
This suggests that the discrepancy between two- and three-pion Gaussian radii are indeed caused by non-Gaussian features of the correlation function.
Concerning the effective intercepts, we observe a substantial increase as compared to the Gaussian case.  

The qualities of the Gaussian, Edgeworth, and exponential fits for three-pion cumulant correlations 
vary depending on the multiplicity interval.  
The $\chi^2/{\rm NDF}$ for the 3D three-pion Gaussian, Edgeworth, and exponential fits in the highest multiplicity \PbPb\ interval is 8600/1436, 4450/1436, and 4030/1436, respectively. 
The $\chi^2/{\rm NDF}$ decreases significantly for lower multiplicity intervals to about 4170/7785 for peripheral \PbPb\ and 12400/17305 
for \pp\ and \pPb\ multiplicity intervals, for all fit types.  
The Edgeworth $\chi^2/{\rm NDF}$ is a few percent smaller than for Gaussian fits in low multiplicity intervals.
The exponential $\chi^2/{\rm NDF}$ is a few percent smaller than for Edgeworth fits in low multiplicity intervals.

Due to the asymmetry of the p--Pb colliding system, the extracted fit parameters in $-0.8<\eta<-0.4$ and $0.4<\eta<0.8$ pseudorapidity intervals are compared.  
The radii and the effective intercept parameters in both intervals are consistent within statistical uncertainties.

The extracted radii in each multiplicity interval and system correspond to different $\Nch$ values.  
To compare the radii in pp and p--Pb at the same $\Nch$ value, we perform a linear fit to the pp three-pion Edgeworth radii as a function of $\Nch^{1/3}$.  
We then compare the extracted p--Pb three-pion Edgeworth radii to the value of the pp fit evaluated at the same $\Nch$.
We find that the Edgeworth radii in p--Pb are on average $10\pm5\%$ larger than for pp in the region of overlapping multiplicity.
The comparison of Pb--Pb to p--Pb radii is done similarly where the fit is performed to p--Pb data and compared to the two-pion Pb--Pb Edgeworth radii.  
The Edgeworth radii in Pb--Pb are found to be on average $45\pm10\%$ larger than for p--Pb in the region of overlapping multiplicity.
The ratio comparison as it is done exploits the cancellation of the largely correlated systematic uncertainties.

\begin{figure}[t]
  \centering
  \subfigure[p--Pb compared to pp]{	
    \includegraphics[width=0.47\textwidth]{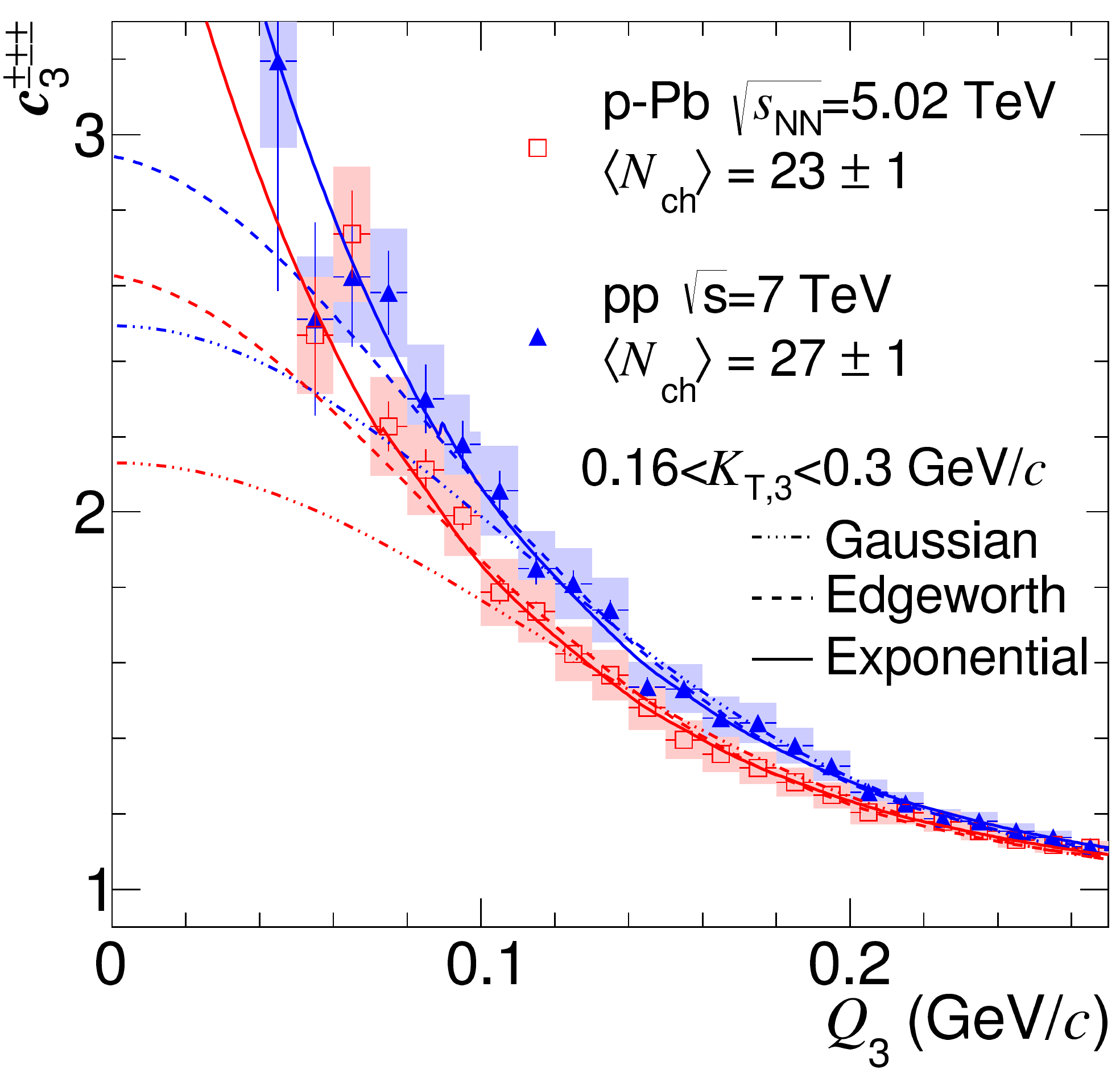}
    \label{fig:pPbAndppComp}	
  } \hspace{0.1cm}
  \subfigure[p--Pb compared to \PbPb\ ]{	
    \includegraphics[width=0.47\textwidth]{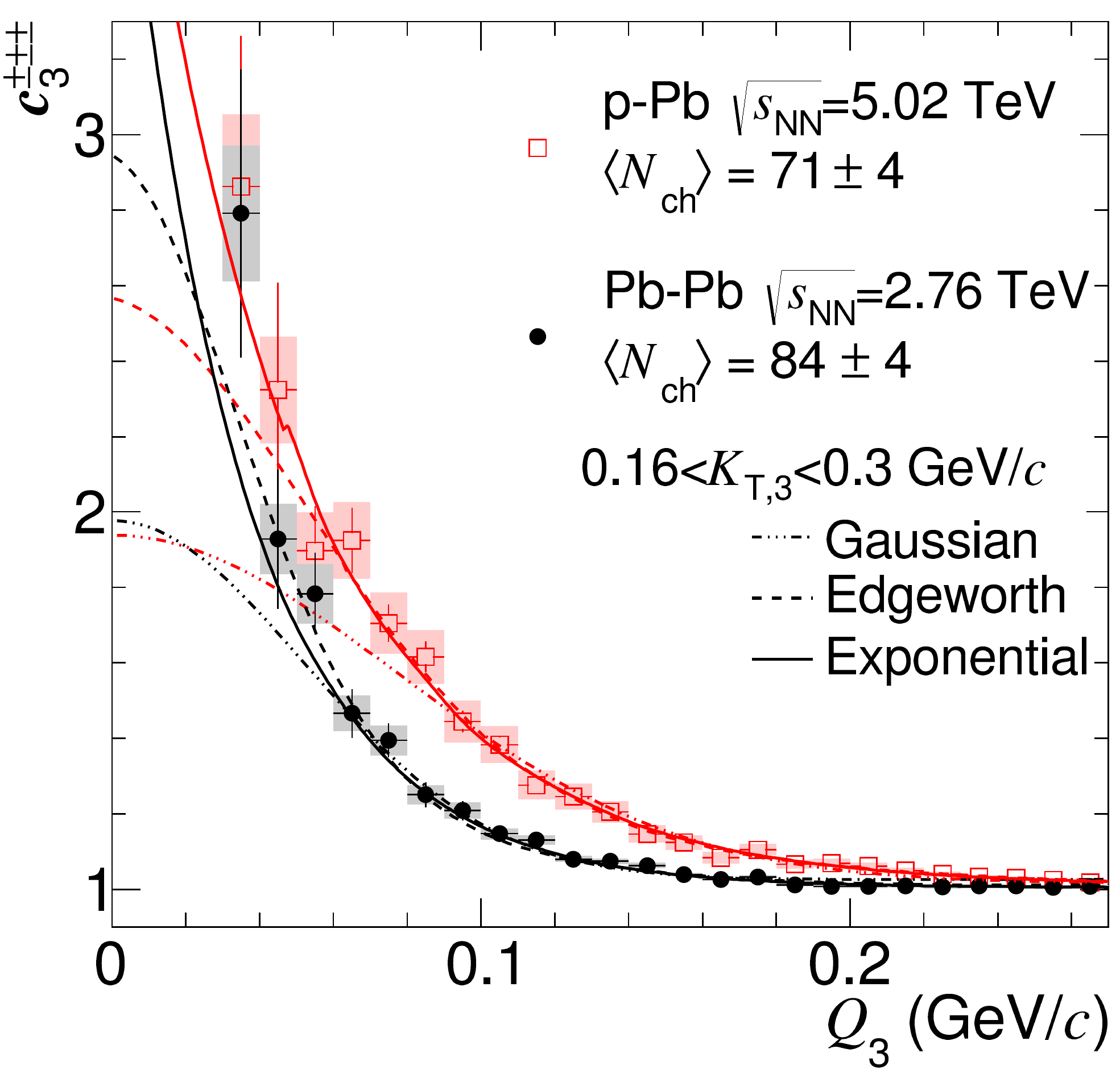}
    \label{fig:pPbAndPbPbComp}	
  }  	
  \caption{Comparisons of same-charge three-pion cumulant correlation functions at similar multiplicity for $0.16<\KTThree<0.3$ GeV/$c$. 
           Three points at low $Q_3$ with large statistical uncertainties are not shown in the left panel.}
\end{figure}
To be independent of the assumed functional form for ${\rm {\bf c}}_3$, the same-charge three-pion cumulant correlation functions are directly compared 
between two collision systems at similar multiplicity. 
Fig.~\ref{fig:pPbAndppComp} shows that while the three-pion correlation functions in \pp\ and \pPb~collisions are different, 
their characteristic widths are similar. It is therefore the $\lst$ values which differ the most between the two systems. 
Fig.~\ref{fig:pPbAndPbPbComp} shows that the correlation functions in \pPb\ and \PbPb~collisions are generally quite different.  

\section{Summary}
\label{sec:summary}
Three-pion correlations of same- and mixed-charge pions have been presented for \pp~($\sqrt{s}=7$ TeV), \pPb~($\snn=5.02$ TeV) and 
\PbPb~($\snn=2.76$ TeV) collisions at the LHC, measured with ALICE.  
Freeze-out radii using Gaussian, Edgeworth, and exponential fits have been extracted from the three-pion cumulant QS correlation 
and presented in intervals of multiplicity and triplet momentum.  
Compared to the radii from two-pion correlations, the radii from three-pion cumulant correlations are less susceptible 
to non-femtoscopic background correlations due to the increased QS signal and the removal of two-pion backgrounds.

The deviation of Gaussian fits below the observed correlations at low $Q_3$ clearly demonstrates the importance of non-Gaussian 
features of the correlation functions.  
The effective intercept parameters from Gaussian (exponential) fits are significantly below (above) the chaotic limits, while the corresponding Edgeworth effective intercepts are much closer to the chaotic limit.

At similar multiplicity, the invariant radii extracted from Edgeworth fits in \pPb\ collisions are found to be 5--15\% larger than those in pp, while those in \PbPb\ are 35--55\% larger than those in p--Pb.
Hence, models which incorporate substantially stronger collective expansion in \pPb\ than \pp\ collisions at similar multiplicity are disfavored.
The comparability of the extracted radii in \pp\ and \pPb\ collisions at similar multiplicity 
is consistent with expectations from CGC initial conditions (IP-Glasma) without a hydrodynamic 
phase~\cite{Bzdak:2013zma}.
The smaller radii in \pPb\ as compared to \PbPb\ collisions may demonstrate the importance of different initial conditions on the final-state, or indicate significant collective expansion already in peripheral \PbPb\ collisions.

\ifpreprint
\iffull
\newenvironment{acknowledgement}{\relax}{\relax}
\begin{acknowledgement}
\section*{Acknowledgements}
We would like to thank Richard Lednick{\' y}, M{\' a}t{\' e} Csan{\' a}d, and Tam{\' a}s Cs{\" o}rg{\H o} for numerous helpful discussions.
\\
The ALICE Collaboration would like to thank all its engineers and technicians for their invaluable contributions to the construction of the experiment and the CERN accelerator teams for the outstanding performance of the LHC complex.
\\
The ALICE Collaboration gratefully acknowledges the resources and support provided by all Grid centres and the Worldwide LHC Computing Grid (WLCG) collaboration.
\\
The ALICE Collaboration acknowledges the following funding agencies for their support in building and
running the ALICE detector:
State Committee of Science,  World Federation of Scientists (WFS)
and Swiss Fonds Kidagan, Armenia,
Conselho Nacional de Desenvolvimento Cient\'{\i}fico e Tecnol\'{o}gico (CNPq), Financiadora de Estudos e Projetos (FINEP),
Funda\c{c}\~{a}o de Amparo \`{a} Pesquisa do Estado de S\~{a}o Paulo (FAPESP);
National Natural Science Foundation of China (NSFC), the Chinese Ministry of Education (CMOE)
and the Ministry of Science and Technology of China (MSTC);
Ministry of Education and Youth of the Czech Republic;
Danish Natural Science Research Council, the Carlsberg Foundation and the Danish National Research Foundation;
The European Research Council under the European Community's Seventh Framework Programme;
Helsinki Institute of Physics and the Academy of Finland;
French CNRS-IN2P3, the `Region Pays de Loire', `Region Alsace', `Region Auvergne' and CEA, France;
German BMBF and the Helmholtz Association;
General Secretariat for Research and Technology, Ministry of
Development, Greece;
Hungarian OTKA and National Office for Research and Technology (NKTH);
Department of Atomic Energy and Department of Science and Technology of the Government of India;
Istituto Nazionale di Fisica Nucleare (INFN) and Centro Fermi -
Museo Storico della Fisica e Centro Studi e Ricerche "Enrico
Fermi", Italy;
MEXT Grant-in-Aid for Specially Promoted Research, Ja\-pan;
Joint Institute for Nuclear Research, Dubna;
National Research Foundation of Korea (NRF);
CONACYT, DGAPA, M\'{e}xico, ALFA-EC and the EPLANET Program
(European Particle Physics Latin American Network)
Stichting voor Fundamenteel Onderzoek der Materie (FOM) and the Nederlandse Organisatie voor Wetenschappelijk Onderzoek (NWO), Netherlands;
Research Council of Norway (NFR);
National Science Centre, Poland;
 Ministry of National Education/Institute for Atomic Physics and CNCS-UEFISCDI - Romania;
Ministry of Education and Science of Russian Federation, Russian
Academy of Sciences, Russian Federal Agency of Atomic Energy,
Russian Federal Agency for Science and Innovations and The Russian
Foundation for Basic Research;
Ministry of Education of Slovakia;
Department of Science and Technology, South Africa;
CIEMAT, EELA, Ministerio de Econom\'{i}a y Competitividad (MINECO) of Spain, Xunta de Galicia (Conseller\'{\i}a de Educaci\'{o}n),
CEA\-DEN, Cubaenerg\'{\i}a, Cuba, and IAEA (International Atomic Energy Agency);
Swedish Research Council (VR) and Knut $\&$ Alice Wallenberg
Foundation (KAW);
Ukraine Ministry of Education and Science;
United Kingdom Science and Technology Facilities Council (STFC);
The United States Department of Energy, the United States National
Science Foundation, the State of Texas, and the State of Ohio.
\end{acknowledgement}
\ifbibtex
\bibliographystyle{h-elsevier}
\bibliography{biblio}{}
\else

\fi
\newpage
\appendix
\ifsupp
\section{Supplemental figures}
\label{app:supp}
\input{supp.tex}
\clearpage
\fi
\section{The ALICE Collaboration}
\label{app:collab}



\begingroup
\small
\begin{flushleft}
B.~Abelev\Irefn{org69}\And
J.~Adam\Irefn{org37}\And
D.~Adamov\'{a}\Irefn{org77}\And
M.M.~Aggarwal\Irefn{org81}\And
M.~Agnello\Irefn{org105}\textsuperscript{,}\Irefn{org88}\And
A.~Agostinelli\Irefn{org26}\And
N.~Agrawal\Irefn{org44}\And
Z.~Ahammed\Irefn{org124}\And
N.~Ahmad\Irefn{org18}\And
I.~Ahmed\Irefn{org15}
% ...
\renewcommand\labelenumi{\textsuperscript{\theenumi}~}

\section*{Affiliation notes}
\renewcommand\theenumi{\roman{enumi}}
\begin{Authlist}
\item \Adef{0}Deceased
\item \Adef{idp1104336}{Also at: St. Petersburg State Polytechnical University}
\item \Adef{idp2979216}{Also at: Department of Applied Physics, Aligarh Muslim University, Aligarh, India}
\item \Adef{idp3656640}{Also at: M.V. Lomonosov Moscow State University, D.V. Skobeltsyn Institute of Nuclear Physics, Moscow, Russia}
\item \Adef{idp3900240}{Also at: University of Belgrade, Faculty of Physics and "Vin\v{c}a" Institute of Nuclear Sciences, Belgrade, Serbia}
\item \Adef{idp4205840}{Permanent Address: Permanent Address: Konkuk University, Seoul, Korea}
\item \Adef{idp4746224}{Also at: Institute of Theoretical Physics, University of Wroclaw, Wroclaw, Poland}
\item \Adef{idp5651280}{Also at: University of Kansas, Lawrence, KS, United States}
\end{Authlist}

\section*{Collaboration Institutes}
\renewcommand\theenumi{\arabic{enumi}~}
\begin{Authlist}

\item \Idef{org1}A.I. Alikhanyan National Science Laboratory (Yerevan Physics Institute) Foundation, Yerevan, Armenia
\item \Idef{org2}Benem\'{e}rita Universidad Aut\'{o}noma de Puebla, Puebla, Mexico
\item \Idef{org3}Bogolyubov Institute for Theoretical Physics, Kiev, Ukraine
\item \Idef{org4}Bose Institute, Department of Physics and Centre for Astroparticle Physics and Space Science (CAPSS), Kolkata, India
\item \Idef{org5}Budker Institute for Nuclear Physics, Novosibirsk, Russia
\item \Idef{org6}California Polytechnic State University, San Luis Obispo, CA, United States
\item \Idef{org7}Central China Normal University, Wuhan, China
\item \Idef{org8}Centre de Calcul de l'IN2P3, Villeurbanne, France
\item \Idef{org9}Centro de Aplicaciones Tecnol\'{o}gicas y Desarrollo Nuclear (CEADEN), Havana, Cuba
\item \Idef{org10}Centro de Investigaciones Energ\'{e}ticas Medioambientales y Tecnol\'{o}gicas (CIEMAT), Madrid, Spain
\item \Idef{org11}Centro de Investigaci\'{o}n y de Estudios Avanzados (CINVESTAV), Mexico City and M\'{e}rida, Mexico
\item \Idef{org12}Centro Fermi - Museo Storico della Fisica e Centro Studi e Ricerche ``Enrico Fermi'', Rome, Italy
\item \Idef{org13}Chicago State University, Chicago, USA
\item \Idef{org14}Commissariat \`{a} l'Energie Atomique, IRFU, Saclay, France
\item \Idef{org15}COMSATS Institute of Information Technology (CIIT), Islamabad, Pakistan
\item \Idef{org16}Departamento de F\'{\i}sica de Part\'{\i}culas and IGFAE, Universidad de Santiago de Compostela, Santiago de Compostela, Spain
\item \Idef{org17}Department of Physics and Technology, University of Bergen, Bergen, Norway
\item \Idef{org18}Department of Physics, Aligarh Muslim University, Aligarh, India
\item \Idef{org19}Department of Physics, Ohio State University, Columbus, OH, United States
\item \Idef{org20}Department of Physics, Sejong University, Seoul, South Korea
\item \Idef{org21}Department of Physics, University of Oslo, Oslo, Norway
\item \Idef{org22}Dipartimento di Fisica dell'Universit\`{a} 'La Sapienza' and Sezione INFN Rome, Italy
\item \Idef{org23}Dipartimento di Fisica dell'Universit\`{a} and Sezione INFN, Cagliari, Italy
\item \Idef{org24}Dipartimento di Fisica dell'Universit\`{a} and Sezione INFN, Trieste, Italy
\item \Idef{org25}Dipartimento di Fisica dell'Universit\`{a} and Sezione INFN, Turin, Italy
\item \Idef{org26}Dipartimento di Fisica e Astronomia dell'Universit\`{a} and Sezione INFN, Bologna, Italy
\item \Idef{org27}Dipartimento di Fisica e Astronomia dell'Universit\`{a} and Sezione INFN, Catania, Italy
\item \Idef{org28}Dipartimento di Fisica e Astronomia dell'Universit\`{a} and Sezione INFN, Padova, Italy
\item \Idef{org29}Dipartimento di Fisica `E.R.~Caianiello' dell'Universit\`{a} and Gruppo Collegato INFN, Salerno, Italy
\item \Idef{org30}Dipartimento di Scienze e Innovazione Tecnologica dell'Universit\`{a} del  Piemonte Orientale and Gruppo Collegato INFN, Alessandria, Italy
\item \Idef{org31}Dipartimento Interateneo di Fisica `M.~Merlin' and Sezione INFN, Bari, Italy
\item \Idef{org32}Division of Experimental High Energy Physics, University of Lund, Lund, Sweden
\item \Idef{org33}Eberhard Karls Universit\"{a}t T\"{u}bingen, T\"{u}bingen, Germany
\item \Idef{org34}European Organization for Nuclear Research (CERN), Geneva, Switzerland
\item \Idef{org35}Faculty of Engineering, Bergen University College, Bergen, Norway
\item \Idef{org36}Faculty of Mathematics, Physics and Informatics, Comenius University, Bratislava, Slovakia
\item \Idef{org37}Faculty of Nuclear Sciences and Physical Engineering, Czech Technical University in Prague, Prague, Czech Republic
\item \Idef{org38}Faculty of Science, P.J.~\v{S}af\'{a}rik University, Ko\v{s}ice, Slovakia
\item \Idef{org39}Frankfurt Institute for Advanced Studies, Johann Wolfgang Goethe-Universit\"{a}t Frankfurt, Frankfurt, Germany
\item \Idef{org40}Gangneung-Wonju National University, Gangneung, South Korea
\item \Idef{org41}Gauhati University, Department of Physics, Guwahati, India
\item \Idef{org42}Helsinki Institute of Physics (HIP), Helsinki, Finland
\item \Idef{org43}Hiroshima University, Hiroshima, Japan
\item \Idef{org44}Indian Institute of Technology Bombay (IIT), Mumbai, India
\item \Idef{org45}Indian Institute of Technology Indore, Indore (IITI), India
\item \Idef{org46}Institut de Physique Nucl\'eaire d'Orsay (IPNO), Universit\'e Paris-Sud, CNRS-IN2P3, Orsay, France
\item \Idef{org47}Institut f\"{u}r Informatik, Johann Wolfgang Goethe-Universit\"{a}t Frankfurt, Frankfurt, Germany
\item \Idef{org48}Institut f\"{u}r Kernphysik, Johann Wolfgang Goethe-Universit\"{a}t Frankfurt, Frankfurt, Germany
\item \Idef{org49}Institut f\"{u}r Kernphysik, Westf\"{a}lische Wilhelms-Universit\"{a}t M\"{u}nster, M\"{u}nster, Germany
\item \Idef{org50}Institut Pluridisciplinaire Hubert Curien (IPHC), Universit\'{e} de Strasbourg, CNRS-IN2P3, Strasbourg, France
\item \Idef{org51}Institute for Nuclear Research, Academy of Sciences, Moscow, Russia
\item \Idef{org52}Institute for Subatomic Physics of Utrecht University, Utrecht, Netherlands
\item \Idef{org53}Institute for Theoretical and Experimental Physics, Moscow, Russia
\item \Idef{org54}Institute of Experimental Physics, Slovak Academy of Sciences, Ko\v{s}ice, Slovakia
\item \Idef{org55}Institute of Physics, Academy of Sciences of the Czech Republic, Prague, Czech Republic
\item \Idef{org56}Institute of Physics, Bhubaneswar, India
\item \Idef{org57}Institute of Space Science (ISS), Bucharest, Romania
\item \Idef{org58}Instituto de Ciencias Nucleares, Universidad Nacional Aut\'{o}noma de M\'{e}xico, Mexico City, Mexico
\item \Idef{org59}Instituto de F\'{\i}sica, Universidad Nacional Aut\'{o}noma de M\'{e}xico, Mexico City, Mexico
\item \Idef{org60}iThemba LABS, National Research Foundation, Somerset West, South Africa
\item \Idef{org61}Joint Institute for Nuclear Research (JINR), Dubna, Russia
\item \Idef{org62}Korea Institute of Science and Technology Information, Daejeon, South Korea
\item \Idef{org63}KTO Karatay University, Konya, Turkey
\item \Idef{org64}Laboratoire de Physique Corpusculaire (LPC), Clermont Universit\'{e}, Universit\'{e} Blaise Pascal, CNRS--IN2P3, Clermont-Ferrand, France
\item \Idef{org65}Laboratoire de Physique Subatomique et de Cosmologie, Universit\'{e} Grenoble-Alpes, CNRS-IN2P3, Grenoble, France
\item \Idef{org66}Laboratori Nazionali di Frascati, INFN, Frascati, Italy
\item \Idef{org67}Laboratori Nazionali di Legnaro, INFN, Legnaro, Italy
\item \Idef{org68}Lawrence Berkeley National Laboratory, Berkeley, CA, United States
\item \Idef{org69}Lawrence Livermore National Laboratory, Livermore, CA, United States
\item \Idef{org70}Moscow Engineering Physics Institute, Moscow, Russia
\item \Idef{org71}National Centre for Nuclear Studies, Warsaw, Poland
\item \Idef{org72}National Institute for Physics and Nuclear Engineering, Bucharest, Romania
\item \Idef{org73}National Institute of Science Education and Research, Bhubaneswar, India
\item \Idef{org74}Niels Bohr Institute, University of Copenhagen, Copenhagen, Denmark
\item \Idef{org75}Nikhef, National Institute for Subatomic Physics, Amsterdam, Netherlands
\item \Idef{org76}Nuclear Physics Group, STFC Daresbury Laboratory, Daresbury, United Kingdom
\item \Idef{org77}Nuclear Physics Institute, Academy of Sciences of the Czech Republic, \v{R}e\v{z} u Prahy, Czech Republic
\item \Idef{org78}Oak Ridge National Laboratory, Oak Ridge, TN, United States
\item \Idef{org79}Petersburg Nuclear Physics Institute, Gatchina, Russia
\item \Idef{org80}Physics Department, Creighton University, Omaha, NE, United States
\item \Idef{org81}Physics Department, Panjab University, Chandigarh, India
\item \Idef{org82}Physics Department, University of Athens, Athens, Greece
\item \Idef{org83}Physics Department, University of Cape Town, Cape Town, South Africa
\item \Idef{org84}Physics Department, University of Jammu, Jammu, India
\item \Idef{org85}Physics Department, University of Rajasthan, Jaipur, India
\item \Idef{org86}Physik Department, Technische Universit\"{a}t M\"{u}nchen, Munich, Germany
\item \Idef{org87}Physikalisches Institut, Ruprecht-Karls-Universit\"{a}t Heidelberg, Heidelberg, Germany
\item \Idef{org88}Politecnico di Torino, Turin, Italy
\item \Idef{org89}Purdue University, West Lafayette, IN, United States
\item \Idef{org90}Pusan National University, Pusan, South Korea
\item \Idef{org91}Research Division and ExtreMe Matter Institute EMMI, GSI Helmholtzzentrum f\"ur Schwerionenforschung, Darmstadt, Germany
\item \Idef{org92}Rudjer Bo\v{s}kovi\'{c} Institute, Zagreb, Croatia
\item \Idef{org93}Russian Federal Nuclear Center (VNIIEF), Sarov, Russia
\item \Idef{org94}Russian Research Centre Kurchatov Institute, Moscow, Russia
\item \Idef{org95}Saha Institute of Nuclear Physics, Kolkata, India
\item \Idef{org96}School of Physics and Astronomy, University of Birmingham, Birmingham, United Kingdom
\item \Idef{org97}Secci\'{o}n F\'{\i}sica, Departamento de Ciencias, Pontificia Universidad Cat\'{o}lica del Per\'{u}, Lima, Peru
\item \Idef{org98}Sezione INFN, Bari, Italy
\item \Idef{org99}Sezione INFN, Bologna, Italy
\item \Idef{org100}Sezione INFN, Cagliari, Italy
\item \Idef{org101}Sezione INFN, Catania, Italy
\item \Idef{org102}Sezione INFN, Padova, Italy
\item \Idef{org103}Sezione INFN, Rome, Italy
\item \Idef{org104}Sezione INFN, Trieste, Italy
\item \Idef{org105}Sezione INFN, Turin, Italy
\item \Idef{org106}SSC IHEP of NRC Kurchatov institute, Protvino, Russia
\item \Idef{org107}SUBATECH, Ecole des Mines de Nantes, Universit\'{e} de Nantes, CNRS-IN2P3, Nantes, France
\item \Idef{org108}Suranaree University of Technology, Nakhon Ratchasima, Thailand
\item \Idef{org109}Technical University of Split FESB, Split, Croatia
\item \Idef{org110}The Henryk Niewodniczanski Institute of Nuclear Physics, Polish Academy of Sciences, Cracow, Poland
\item \Idef{org111}The University of Texas at Austin, Physics Department, Austin, TX, USA
\item \Idef{org112}Universidad Aut\'{o}noma de Sinaloa, Culiac\'{a}n, Mexico
\item \Idef{org113}Universidade de S\~{a}o Paulo (USP), S\~{a}o Paulo, Brazil
\item \Idef{org114}Universidade Estadual de Campinas (UNICAMP), Campinas, Brazil
\item \Idef{org115}University of Houston, Houston, TX, United States
\item \Idef{org116}University of Jyv\"{a}skyl\"{a}, Jyv\"{a}skyl\"{a}, Finland
\item \Idef{org117}University of Liverpool, Liverpool, United Kingdom
\item \Idef{org118}University of Tennessee, Knoxville, TN, United States
\item \Idef{org119}University of Tokyo, Tokyo, Japan
\item \Idef{org120}University of Tsukuba, Tsukuba, Japan
\item \Idef{org121}University of Zagreb, Zagreb, Croatia
\item \Idef{org122}Universit\'{e} de Lyon, Universit\'{e} Lyon 1, CNRS/IN2P3, IPN-Lyon, Villeurbanne, France
\item \Idef{org123}V.~Fock Institute for Physics, St. Petersburg State University, St. Petersburg, Russia
\item \Idef{org124}Variable Energy Cyclotron Centre, Kolkata, India
\item \Idef{org125}Vestfold University College, Tonsberg, Norway
\item \Idef{org126}Warsaw University of Technology, Warsaw, Poland
\item \Idef{org127}Wayne State University, Detroit, MI, United States
\item \Idef{org128}Wigner Research Centre for Physics, Hungarian Academy of Sciences, Budapest, Hungary
\item \Idef{org129}Yale University, New Haven, CT, United States
\item \Idef{org130}Yonsei University, Seoul, South Korea
\item \Idef{org131}Zentrum f\"{u}r Technologietransfer und Telekommunikation (ZTT), Fachhochschule Worms, Worms, Germany
\end{Authlist}
\endgroup

\else
\ifbibtex
\bibliographystyle{h-elsevier}
\bibliography{biblio}{}
\else

\fi
\fi
\else
\iffull
\vspace{0.5cm}

\input{refpaper.tex}
\else
\ifbibtex
\bibliographystyle{h-elsevier-notitle}
\bibliography{biblio}{}
\else
\input{refpaper.tex}
\fi
\fi
\fi
\end{document}